\documentclass[12pt]{article}

\usepackage{amsmath}
\usepackage{multirow}
\usepackage{amsfonts}
\usepackage{amssymb}
\def\im{\mathrm{i}}

\def\hybrid{\topmargin -20pt
\oddsidemargin 0pt
\headheight 0pt
\headsep 0pt
\textwidth 6.25in
\textheight 9.5in
\marginparwidth .875in
\parskip 5pt plus 1pt
\jot = 1.5ex}
\hybrid
\numberwithin{equation}{section}
\numberwithin{table}{section}
\begin{document}

\begin{titlepage}

\begin{center}

\rightline{\small ZMP-HH/08-2}

\vskip 2.5cm

{\large \bf The effective action of the heterotic string compactified on manifolds with $SU(3)$ structure}
\footnote{Work supported by: DFG -- The German Science Foundation, European RTN Program MRTN-CT-2004-503369 and the DAAD -- the German Academic Exchange Service.}\\

\vskip 2cm

{\bf Iman Benmachiche, Jan Louis$^{a,b}$ and Danny Mart\'\i nez-Pedrera$^{a}$}\\

\vskip 1.5cm

{}$^{a}${\em II. Institut f{\"u}r Theoretische Physik\\
Universit{\"a}t Hamburg\\
Luruper Chaussee 149\\
D-22761 Hamburg, Germany}\\

\vskip 0.4cm

{}$^{b}${\em Zentrum f\"ur Mathematische Physik, Universit\"at Hamburg,\\
Bundesstrasse 55, D-20146 Hamburg}

\vskip 1cm 

{\tt i.benmachiche@gmail.com, jan.louis@desy.de, danny@mail.desy.de} \\

\end{center}

\vskip 2cm

\begin{center} {\bf ABSTRACT } \end{center}

We derive the $N=1$ effective action of the heterotic string compactified on manifolds with $SU(3)$ structure in the presence of background fluxes. We use a Kaluza-Klein reduction and compute the moduli dependence of the K\"ahler potential, the gauge kinetic function and the superpotential entirely from fermionic terms of the reduced action.

\vfill

February 2008

\end{titlepage}


\section{Introduction}

Heterotic string theory is a promising candidate to incorporate the supersymmetric standard model of particle physics. It naturally includes chiral fermions in non-Abelian representations of a gauge group which is big enough to comfortably accommodate the $\mathrm{SU}(3)\times\mathrm{SU}(2)\times\mathrm{U}(1)$ of the Standard Model. As a consequence, string phenomenology has focused for many years on the study of $N=1$ heterotic string vacua.\footnote{For reviews see, for example, \cite{GR,Quevedo:1997uy}; more recent development can be found, for example, in \cite{Braun:2005ux,Buchmuller:2006ik,Blumenhagen:2006ux}.}

Heterotic string vacua with an unbroken $N=1$ supersymmetry in four space-time dimensions can be constructed either by compactifying the ten-dimensional heterotic string on six-dimensional Calabi-Yau threefolds $Y$ or by considering the heterotic string with an appropriate internal $(0,2)$ superconformal field theory \cite{Greene:1996cy}.

One of the problems that one faces in following this program is the identification of a mechanism for spontaneous supersymmetry breaking at a scale hierarchically lower than the Planck scale. Traditionally, one employs non-perturbative effects such as gaugino condensation and includes them at the level of the low energy effective action \cite{Dine:1985rz,VY,KL}. A second problem is the vacuum degeneracy of string theory which is parametrized by the vacuum expectation values (VEVs) of gauge-neutral moduli scalars. As a consequence, phenomenological predictions of string theory are difficult to extract. However, depending on the mechanism of supersymmetry breaking, the scalar field VEVs might be fixed by the same potential responsible for the breaking of supersymmetry.

An alternative mechanism where supersymmetry is broken by non-trivial background fluxes of the NS three-form $H_3$ was proposed in references \cite{Rohm:1985jv,Strominger:1986uh,deWit:1986xg} and has been revived in \cite{Bachas:1995ik}--\cite{Louis:2001uy}. Similarly, generalized compactifications on manifolds $\hat Y$ with $\mathrm{SU}(3)$ structure (rather than $\mathrm{SU}(3)$ holonomy) have been considered \cite{Rocek:1991ze}--\cite{Becker:2003yv}.\footnote{For reviews and a more complete list of references see, for example, \cite{grana}.} These manifolds admit a globally defined spinor $\eta$  but contrary to Calabi-Yau manifolds this spinor is no longer covariantly constant with respect to the Levi-Civita connection. Instead, $\hat Y$ has torsion, and depending on the fluxes the supersymmetry can be spontaneously broken. The fluxes as well as the torsion induce a non-trivial superpotential $W$ in the effective action and hence fix at least part of the moduli scalars. When combined with non-perturbative effects they might lead to viable scenarios of hierarchical supersymmetry breaking.

The computation of the effective action on manifolds with $\mathrm{SU}(3)$ structure is not entirely straightforward. The reason is that the distinction between heavy and light modes is not as clean as it is for Calabi-Yau manifolds $Y$, where the massless modes are in one-to-one correspondence with the harmonic forms on $Y$ and consequently one only keeps these in the Kaluza-Klein reduction. For manifolds with $\mathrm{SU}(3)$ structure a similar characterization is missing so far. Here we follow the approach outlined in \cite{Grana:2005ny} (which is based on \cite{Gurrieri:2002wz}) where type II supergravities are studied in backgrounds which break the ten-dimensional Lorentz symmetry $\mathrm{SO}(1,9)$ to $\mathrm{SO}(1,3)\times\mathrm{SU}(3)$.\footnote{This approach has been refined in refs.~\cite{KashaniPoor:2006si,Koerber:2007xk}.} In such backgrounds one can already in ten dimensions and without performing a Kaluza-Klein reduction rewrite the type II action in a form which resembles a four-dimensional $N=2$ action. More precisely, one finds (and computes) the typical couplings of an $N=2$ action such as the special K\"ahler potentials and the Killing prepotentials. In a second step one can then truncate this ten-dimensional action to a four-dimensional effective action keeping only a finite subset of Kaluza-Klein modes. In order to ensure a consistent four-dimensional $N=2$ effective action one needs to perform the truncation in such a way that the $N=2$ couplings of the ten-dimensional action properly descend to the four-dimensional action. Furthermore, projecting out all $\mathrm{SU}(3)$ triplets results in a `standard' $N=2$ action without any massive gravitino multiplet.

Here we follow a similar strategy for the heterotic string. However, we will not be as explicit as in \cite{Grana:2005ny} and directly perform the Kaluza-Klein reduction keeping a finite subset of light modes. Among the light modes we do not keep any triplets of $\mathrm{SU}(3)$ or, in other words, we do not keep light modes which arise from one-forms (or five-forms) of $\hat Y$. Apart from this constraint, we keep the manifold $\hat Y$ generic.

Various properties of the couplings of the $N=1$ effective action have been computed previously in \cite{Cardoso:2002hd,Becker:2003yv} and \cite{Becker:2003gq}--\cite{Ali:2007ra}. The purpose of this paper is to obtain the couplings for a generic $\hat Y$ with background fluxes entirely from fermionic terms in the low energy action. This has the advantage that the superpotential $W$ and its derivatives appear linearly (as opposed to quadratically in the bosonic terms) and therefore can be computed straightforwardly \cite{Gurrieri:2004dt,Jockers:2005zy,Benmachiche:2006df}. However, in order to do so one also needs the proper normalization of the fermionic kinetic terms. To our surprise and as far as we know, the fermionic terms have not been written down even for Calabi-Yau compactifications.

The paper is organized as follows. We start by reviewing some basic facts about $\mathrm{SU}(3)$-structure manifolds in section~\ref{sec-man-su3}. In section~\ref{sec-4d-spec} we derive the four-dimensional $N=1$ spectrum via Kaluza-Klein reduction. Here we leave the structure of the gauge bundle somewhat arbitrary or in other words we do not explicitly solve the Bianchi identity for the three-form field strength $H$. However, whenever necessary we discuss the case of the standard embedding as an instructuve example \cite{Gurrieri:2007jg,Ali:2007ra}. The $N=1$ low energy effective action for compactifications on generalized manifolds $\hat Y$ with $\mathrm{SU}(3)$ structure and background flux is  determined in section~\ref{sec-eff-act}. More specifically, in section \ref{bosonicK} we recall the bosonic kinetic terms. In section \ref{fermionicK} we compute the kinetic terms of all fermionic fields, which includes as a special case also Calabi-Yau compactifications. From these terms the K\"ahler potential and the gauge kinetic function are deduced, confirming the structure of these couplings derived previously from the bosonic terms \cite{Strominger:1985ks,Dixon:1989fj,Candelas:1990pi}. For us the  necessity of this step is that it fixes the correct normalization of the fermions, which we need in order to reliably compute the superpotential induced by the presence of non-trivial background fluxes and torsion. It is worth remarking that this computation is not confined to the heterotic string but does determine the kinetic terms in the NS sector for all string theories. In section~\ref{Yukawa} we compute the Yukawa couplings of the matter fields while in section~\ref{Fterms} we determine the dependence of the superpotential on the background fluxes and torsion from the gravitino mass and other fermionic couplings. We compute the $D$-terms in section \ref{sec-dterm}. Finally in section~\ref{sec-susy} we derive the supersymmetry transformations for the four-dimensional gravitino and the chiral fermions, which further checks the validity of the obtained expression for the superpotential. This will allow us to discuss also the conditions for a supersymmetric vacuum, consistent with Strominger's results \cite{Strominger:1986uh}.

A preliminary version of our work appeared in \cite{Imanthesis}. While this manuscript was being prepared the paper \cite{Gurrieri:2007jg} appeared which has some overlap with our work.

\section{Manifolds with $\mathrm{SU}(3)$ structure}\label{sec-man-su3}

In this paper we consider the heterotic string in a space-time background of the form
\begin{equation}
M_4\times_\mathrm{w}\hat Y\ ,
\label{backg}
\end{equation}
where $M_4$ is a four-dimensional Minkowski space and $\hat Y$ is a compact six-dimensional manifold. As it is indicated, the product structure can be warped, which means that the ten-dimensional metric is block diagonal and reads
\begin{equation}
ds^2 = e^{2\Delta(y)} g_{\mu\nu} dx^\mu dx^\nu + g_{mn} dy^m dy^n\ ,
\end{equation}
where $x^\mu$, $\mu = 0,\ldots,3$ are the coordinates of $M_4$ while $y^m$, $m=1,\ldots,6$ are the coordinates of $\hat Y$. For most parts of this paper we perform the analysis in a regime where the warp factor $\Delta$ can be neglected.

Furthermore, we require that in the background (\ref{backg}) a four-dimensional $N=1$ supersymmetry exists. This in turn demands a reduction of the structure group $\mathrm{SO}(6)$ of $\hat Y$ such that a single spinor $\eta$ is globally defined on $\hat Y$. Manifolds with this property are termed `manifolds with $\mathrm{SU}(3)$ structure' \cite{salamonb}. Under $\mathrm{SU}(3)$, the spinor representation $\mathbf{4}$ of $\mathrm{SO}(6)$ decomposes as $\mathbf{4} \to \mathbf{3} + \mathbf{1}$, with $\eta$ being the singlet $\mathbf{1}$.

Calabi-Yau manifolds are a special class of manifolds with $\mathrm{SU}(3)$ structure where $\eta$ is covariantly constant with respect to the Levi-Civita connection or in other words the holonomy of $\hat Y$ is $\mathrm{SU}(3)$. In general $\eta$ is only parallel with respect to a different connection which has torsion, i.e. $\nabla^{(T)}\eta=0$. The part of the torsion which is independent of the choice of $\nabla^{(T)}$ is known as the `intrinsic torsion' and can be used to classify the different types of $\mathrm{SU}(3)$ structures \cite{CS}.

Once $\eta$ exists one can use it to define a two-form $J$ and a three-form $\Omega_\eta$ as follows\footnote{We use this notation for the three-form in order to emphasize the fact that it is constructed from a normalized spinor $\eta$ and to distinguish it from a three-form $\Omega$ to be introduced later which obeys a different normalization.}
\begin{equation}
J^{mn} = \mp \im \eta_\pm^\dag \gamma^{mn} \eta_\pm\ ,\qquad
\Omega_\eta^{mnp} = -\im \eta_-^\dag \gamma^{mnp} \eta_+\ ,
\label{jomega}
\end{equation}
where $\eta_\pm$ is a positive (negative) chirality spinor normalized as $\eta_\pm^\dag \eta_\pm = 1$. $\gamma^{m_1\ldots m_p} = \frac1{n!}\gamma^{[m_1}\gamma^{m_2}\ldots \gamma^{m_p]}$ are antisymmetrized products of six-dimensional $\gamma$-matrices.\footnote{For more details on our conventions see appendix \ref{app-conv}.} Using appropriate Fierz identities one shows that with this normalization for the spinors, $J$ and $\Omega_\eta$ are not independent but satisfy
\begin{equation}
J\wedge J\wedge J = \tfrac34 \im\, \Omega_\eta \wedge \bar\Omega_\eta\ ,\qquad J\wedge \Omega_\eta = 0\ .
\end{equation}
Furthermore, using $\gamma$-matrix identities it can be shown that $J^n_m$ (the index is raised with the metric $g_{mn}$) defines an almost complex structure in that $J^n_m J^p_n = - \delta^p_m$ holds. With respect to this almost complex structure $J_{mn}$ is a $(1,1)$-form while $\Omega_\eta$ is a $(3,0)$-form.

Neither $J$ nor $\Omega_\eta$ is generically close. Instead, $dJ$ and $d\Omega_\eta$ define  the five torsion classes $\mathcal{W}_1, \ldots, \mathcal{W}_5$. Explicitly one has \cite{CS}
\begin{equation}
\begin{aligned}
dJ &= - \frac32 \mathrm{Im} (\mathcal{W}_1 \bar\Omega_\eta) + \mathcal{W}_4 \wedge J + \mathcal{W}_3\ ,\\
d\Omega_\eta &= \mathcal{W}_1 J\wedge J + \mathcal{W}_2 \wedge J + \overline{\mathcal{W}}_5 \wedge \Omega_\eta\ ,
\end{aligned}
\label{djdomega}
\end{equation}
with
\begin{equation}
\mathcal{W}_3\wedge J = \mathcal{W}_3 \wedge \Omega_\eta = \mathcal{W}_2 \wedge J \wedge J = 0\ .
\end{equation}
These five torsion classes completely determine the intrinsic torsion. Note that $\mathcal{W}_1$ is a zero-form, $\mathcal{W}_4$ and $\mathcal{W}_5$ are one-forms, $\mathcal{W}_2$ is a two-form and $\mathcal{W}_3$ is a three-form, and each one can be characterized by the $\mathrm{SU}(3)$ transformation properties. Calabi-Yau manifolds are manifolds of $\mathrm{SU}(3)$ structure where all five torsion classes vanish. Any subset of vanishing torsion classes on the other hand defines specific classes of $\mathrm{SU}(3)$ structure manifolds. For example, projecting out all triplets amounts to setting $\mathcal{W}_4 = \mathcal{W}_5 = 0$ which is the case we consider in this paper.

\section{The four-dimensional spectrum}\label{sec-4d-spec}

Our next task is to compute the light four-dimensional spectrum in the background (\ref{backg}) by a Kaluza-Klein reduction. We start from the ten-dimensional $N=1$ supergravity, which contains a gravitational multiplet consisting of the ten-dimensional metric $\hat G_{MN}, M, N= 0,\ldots 9$, an antisymmetric two-tensor $\hat B_{MN}$, the dilaton $\hat\phi$, a left-handed Majorana-Weyl gravitino $\hat\psi_M$ and a right handed Majorana-Weyl fermion, the dilatino $\hat\lambda$. Additionally, we have a Yang-Mills vector multiplet which features a gauge boson $\hat A^{\hat A}_M$ and a gaugino $\hat\chi^{\hat A}$, both transforming in the adjoint representation of either $E_8\times E_8$ or $\mathrm{SO}(32)$.\footnote{We denote ten-dimensional fields, as well as indices in the adjoint of the ten-dimensional gauge group, by a `hat', e.g. $\hat\chi^{\hat A}$.} We summarize the ten-dimensional spectrum in table \ref{tab-10d-spec}.

\begin{table}[h]
\begin{center}
\begin{tabular}{|c|c|c|}
\hline
multiplet&bosons&fermions\\
\hline
gravitational&$\hat G_{MN}$, $\hat B_{MN}$, $\hat\phi$&$\hat\psi_M$, $\hat\lambda$\\
\hline
vector&$\hat A^{\hat A}_M$&$\hat\chi^{\hat A}$\\
\hline
\end{tabular}
\caption{\label{tab-10d-spec}$N=1$ spectrum in $D=10$.}
\end{center}
\end{table}

In order to prepare for the Kaluza-Klein reduction in the background (\ref{backg}) let us decompose the ten-dimensional fields into representations of the structure group $\mathrm{SU}(3)$ following \cite{Grana:2005ny}. This decomposition is summarized in table \ref{tab-su3-bos} for the bosons and in table \ref{tab-su3-fer} for the fermions. The notation $\mathbf{a}_\mathbf{b}$ denotes a field in the $\mathrm{SU}(3)$ representation $\mathbf{a}$ with a four-dimensional spin $\mathbf{b}$. \textbf{T} denotes an antisymmetric tensor or pseudo-scalar. Note that the components in the vector multiplet also carry an index in some representation of the gauge group which we suppress in tables \ref{tab-su3-bos} and \ref{tab-su3-fer}.

\begin{table}[ht]
\begin{center}
\begin{tabular}{|c|c|l|}
\hline
\multirow{3}{*}{$\hat G_{MN}$} & $g_{\mu\nu}$ & $\mathbf{1}_\mathbf{2}$ \\ \cline{2-3}
& $g_{\mu m}$ & $(\mathbf{3}+\mathbf{\bar 3})_\mathbf{1}$ \\ \cline{2-3}
& $g_{mn}$ & $\mathbf{1}_\mathbf{0} + (\mathbf{6} + \mathbf{\bar 6})_\mathbf{0} + \mathbf{8}_\mathbf{0}$ \\
\hline
\multirow{3}{*}{$\hat B_{MN}$} & $B_{\mu\nu}$ & $\mathbf{1}_\mathbf{T}$ \\ \cline{2-3}
& $B_{\mu m}$ & $(\mathbf{3}+\mathbf{\bar 3})_\mathbf{1}$ \\ \cline{2-3}
& $B_{mn}$ & $\mathbf{1}_\mathbf{0} + (\mathbf{6} + \mathbf{\bar 6})_\mathbf{0} + \mathbf{8}_\mathbf{0}$ \\
\hline
$\hat\phi$ & $\phi$ & $\mathbf{1}_\mathbf{0}$ \\
\hline
\multirow{2}{*}{$\hat A^{\hat A}_M$} & $A_\mu$ & $\mathbf{1}_\mathbf{1}$ \\ \cline{2-3}
& $A_m$ & $(\mathbf{3}+\mathbf{\bar 3})_\mathbf{0}$ \\
\hline
\end{tabular}
\end{center}
\caption{\label{tab-su3-bos}Decomposition of the NS sector in $\mathrm{SU}(3)$ representations.}
\end{table}

\begin{table}[ht]
\begin{center}
\begin{tabular}{|c|c|l|}
\hline
\multirow{2}{*}{$\hat\psi_M$} & $\psi_\mu$ & $\mathbf{1}_\mathbf{3/2} + \mathbf{3}_\mathbf{3/2}$ \\ \cline{2-3}
& $\psi_m$ & $\mathbf{1}_\mathbf{1/2} + \mathbf{3}_\mathbf{1/2} + 2 \cdot\mathbf{\bar 3}_\mathbf{1/2} + \mathbf{6}_\mathbf{1/2} + \mathbf{8}_\mathbf{1/2}$  \\
\hline
$\hat \lambda$ & $\lambda$ & $\mathbf{1}_\mathbf{1/2} + \mathbf{3}_\mathbf{1/2}$ \\
\hline
$\hat \chi$ & $\chi$ & $\mathbf{1}_\mathbf{1/2} + \mathbf{3}_\mathbf{1/2}$ \\
\hline
\end{tabular}
\end{center}
\caption{\label{tab-su3-fer}Decomposition of the fermions in $\mathrm{SU}(3)$ representations.}
\end{table}

In order to perform the Kaluza-Klein reduction we need to determine the light modes. For Calabi-Yau compactifications this is straightforward and well-established: the light modes are in one-to-one correspondence with the harmonic forms on the Calabi-Yau~\cite{GR}. However, for compactification on generic $\mathrm{SU}(3)$ structure manifolds the distinction between heavy and light is more subtle. Here we adopt the approach followed in \cite{Gurrieri:2002wz,Grana:2005ny} in that we expand in a finite set of forms which are not necessarily harmonic. Since we are interested in a standard $N=1$ effective theory we only keep one gravitino in the gravitational multiplet and project out all other gravitini. By inspecting table \ref{tab-su3-fer} we see that this is ensured by leaving out all modes transforming in the $\mathbf{3}$ or $\mathbf{\bar 3}$ of $\mathrm{SU}(3)$.

For the fields arising from the gauge bosons this procedure is model dependent since one needs to specify the structure of the gauge bundle on $\hat Y$ and solve the Bianchi identity of the three-form
field strength $H$. A detailed discussion of this point is beyond the scope of this paper. However, for concreteness we base our discussion on the standard embedding of Calabi-Yau manifolds where the spin connection is identified with an $\mathrm{SU}(3)$ subgroup of the $E_8$ gauge connection \cite{Gurrieri:2007jg}. In this case, the gauge group is decomposed as
\begin{equation}
E_8 \to \mathrm{SU}(3)\times E_6,
\label{standembed}
\end{equation}
and the $\mathbf{248}$ adjoint representation decomposes accordingly to
\begin{equation}
\mathbf{248} \to (\mathbf{1},\mathbf{78}) \oplus (\mathbf{8},\mathbf{1}) \oplus (\mathbf{3},\mathbf{27}) \oplus (\mathbf{\bar 3},\mathbf{\bar{27}}).
\label{adjdecomp}
\end{equation}
Projecting out the triplets of $\mathrm{SU}(3)$ in $\hat A^{\hat A}_M$ leaves us with a vector field in four dimensions $A^A_\mu$, where the index $A$ labels the adjoint of the unbroken $E_6\times E_8$. From  $A^{\hat A}_m$ we obtain instead scalar fields in the $(\mathbf{8} + \mathbf{1},\mathbf{27})$ and
$(\mathbf{6},\mathbf{\bar{27}})$ of $\mathrm{SU}(3)\times E_6$. Similarly, the ten-dimensional gaugino yields  a gaugino in four-dimensions $\chi^A$ in the adjoint of $E_6\times E_8$ and chiral matter fermions in the $(\mathbf{8} + \mathbf{1},\mathbf{27})$ and $(\mathbf{6},\mathbf{\bar{27}})$ of $\mathrm{SU}(3)\times E_6$.

These fields, together with the ones descending from the gravitational sector in ten-dimensions, arrange themselves in $N = 1$ multiplets as shown in table \ref{tab-mult-struc}. In the general case the vector multiplet will transform in the adjoint of $G\times E_8$, where $G$ will depend on the precise form of the gauge bundle, and the chiral matter multiplets will transform also in appropriate representations of $G$. In most of the following the precise gauge structure will play no role and we will only need the $\mathrm{SU}(3)$ representation. In fact in most of the text the index $A$ denoting the adjoint of the gauge group will not be shown explicitly.

\begin{table}[ht]
\begin{center}
\begin{tabular}{|c|c|c|}
\hline
multiplet & $\mathrm{SU}(3)$ rep. & field content \\ \hline
gravitational & $\mathbf{1}$ & $(g_{\mu\nu}, \psi_\mu)$ \\ \hline
linear & $\mathbf{1}$ & $(B_{\mu\nu}, \lambda)$ \\ \hline
vector & $\mathbf{1}$ & $(A^A_\mu, \chi^A)$ \\ \hline
\multirow{2}{*}{chiral moduli} & $\mathbf{6}$ & $(g_{mn}, \psi_m)$ \\ \cline{2-3}
& $\mathbf{8} + \mathbf{1}$ & $(g_{mn}, B_{mn}, \psi_m)$ \\ \hline
\multirow{2}{*}{chiral matter} & $\mathbf{6}$ & $(A_m, \chi)$ \\ \cline{2-3}
& $\mathbf{8} + \mathbf{1}$ & $(A_m, \chi)$ \\ \hline
\end{tabular}
\end{center}
\caption{\label{tab-mult-struc}$N=1$ multiplets.}
\end{table}

So far we merely decomposed the ten-dimensional fields according to their $\mathrm{SU}(3)$ representations and projected out all the triplets. The next step is to expand these fields in a finite basis of forms on $\hat Y$. For Calabi-Yau manifolds one chooses the harmonic $(p,q)$-forms, and in this way only modes which are massless from a four-dimensional viewpoint are kept. On generic manifolds with $\mathrm{SU}(3)$ structure there always exists an almost complex structure $J$, as we reviewed in the previous section, and one can still define $(p,q)$-forms. Truncating the triplets amounts to keeping states which arise from expanding the ten-dimensional fields in $(1,1)$- and $(1,2)$-forms only. Let us now turn to this expansion in some more detail starting with the bosonic fields in section \ref{sec-spec-bos} and discussing the fermions in section \ref{sec-spec-fer}.

\subsection{Bosonic spectrum}\label{sec-spec-bos}

Since the dilaton $\hat \phi$ is already a scalar in $D=10$ it trivially descends to the four-dimensional theory, $\hat\phi(x,y) = \phi(x)$. From table \ref{tab-su3-bos} we see that for the antisymmetric tensor $\hat B_{MN}$ (or $\hat B_2$) one has two contributions,
\begin{equation}
\hat B_2 = B_2(x) + b^i (x)\, \omega_i\ , \qquad i=1,\ldots,h^{(1,1)}\ ,
\label{b2}
\end{equation}
where $B_2(x)$ is a two-form in $D=4$ while the $b^i$ are $h^{(1,1)}$ four-dimensional scalar fields. The $\omega_i$ are a set of $h^{(1,1)}$ $(1,1)$-forms which transform in the $\mathbf{8} \oplus \mathbf{1}$ of $\mathrm{SU}(3)$ but they  are not necessarily harmonic.\footnote{Here we are using the notation that is usual in Calabi-Yau compactifications, even though the forms are not harmonic.}

The ten-dimensional metric $\hat G_{MN}$ decomposes as shown in table \ref{tab-su3-bos}, with $g_{\mu\nu}$ being the bosonic component of the $N=1$ gravitational multiplet. The deformations of the internal part of the metric $\delta g_{mn}$ give rise to two distinct classes of scalar fields corresponding to the $\mathbf{8} \oplus \mathbf{1}$ and the $\mathbf{6} \oplus \mathbf{\bar 6}$ representations of $\mathrm{SU}(3)$. These are most easily distinguished by going to complex indices $\alpha, \bar\beta = 1,2,3$ with respect to the complex structure $J$. In this notation $\delta g_{\alpha\bar\beta}$ transforms in the $\mathbf{8} \oplus \mathbf{1}$ representation while $\delta g_{\alpha\beta}$ transforms in the $\mathbf{6}$. Both deformations are expanded in an appropriate basis of forms on $\hat Y$ as follows
\begin{equation}
\begin{aligned}
\delta g_{\alpha\bar\beta} &= -\im\, \tilde v^i(x) (\omega_i)_{\alpha\bar\beta}\ , \qquad \alpha,\bar\beta = 1,2,3\ , \qquad i = 1, \ldots, h^{(1,1)}\ ,\\
\delta g_{\alpha\beta} &= \frac\im{\Vert\Omega\Vert^2}\, \bar z^a(x) (\bar\rho_a)_{\alpha\bar\gamma\bar\delta}\, {\Omega_\beta}^{\bar\gamma\bar\delta}\ , \qquad a = 1,\ldots,h^{(1,2)}\ ,
\label{deltag}
\end{aligned}
\end{equation}
where the $\omega_i$ are the $(1,1)$-forms already used in (\ref{b2}) while the $\rho_a$ are a set of $h^{(1,2)}$ $(1,2)$-forms transforming in the $\mathbf{6}$ of $\mathrm{SU}(3)$. $\Omega$ is the $(3,0)$-form on $\hat Y$ which differs from the $\Omega_\eta$ introduced in (\ref{jomega}) by a factor $\Omega = \Vert\Omega\Vert \Omega_\eta$ with $\Vert\Omega\Vert^2 \equiv \frac1{3!} \Omega_{\alpha\beta\gamma} \bar\Omega^{\alpha\beta\gamma}$. $\Vert\Omega\Vert$ is constant on the manifold but as reviewed in appendix \ref{app-geom} does depend on the scalar fields. It is introduced in (\ref{deltag}) for later convenience to ensure a properly normalized metric on the space of metric deformations \cite{Candelas:1990pi}. The coefficients in the expansion (\ref{deltag}) correspond to scalar fields on $M_4$. More specifically, the $\tilde v^i(x)$ are $h^{(1,1)}$ real scalars while $z^a(x)$ are $h^{(1,2)}$ complex scalar fields.\footnote{In Calabi-Yau compactifications the $\tilde v^i$ are deformations of the K\"ahler form while the $z^a$ are deformations of the complex structure (for more details see appendix \ref{app-geom}).} The expansion given in (\ref{deltag}) features the metric in the Einstein frame but as will be seen in the next section the correct four-dimensional field variables $v^i$ arise from the expansion of the metric in the string frame. The two metrics differ by a dilaton-dependent factor which relates the scalar fields as follows
\begin{equation}
\tilde v^i = v^i e^{-\phi/2}\ .
\label{frames}
\end{equation}
The $v^i$ combine with the $b^i$ introduced in (\ref{b2}) to form complex scalars $t^i = b^i + \im v^i$ as anticipated in table \ref{tab-mult-struc}.

As already discussed the ten-dimensional gauge field gives rise to the four-dimensional gauge fields $A^A_\mu$ which are singlets under $\mathrm{SU}(3)$ and transform in the adjoint representation of $G\times E_8$. In addition, charged scalar fields arise from $A_m$ which, as we learned from the $\mathrm{SU}(3)$ group theory decomposition of the previous section, sit either in the $\mathbf{6}$ or in the $\mathbf{8} \oplus \mathbf{1}$ of $\mathrm{SU}(3)$ and transform in some representations of $G$ (see table \ref{tab-mult-struc}). Hence we decompose
\begin{equation}
\hat A_{\alpha\beta} = \frac1{\Vert\Omega\Vert^2} A^a(x) (\bar\rho_a)_{\alpha\bar\gamma\bar\delta}
{\Omega_\beta}^{\bar\gamma\bar\delta}\ , \qquad \hat A_{\alpha\bar\beta} = A^i(x) (\omega_i)_{\alpha\bar\beta}\ ,
\label{adecomp}
\end{equation}
where in each case an index labeling the representation of the gauge group under which these fields transform is suppressed. We summarize the four-dimensional spectrum in table \ref{tab-4d-spec}.

\begin{table}[ht]
\begin{center}
\begin{tabular}{|c|c|c|c|}
\hline
multiplet & multiplicity & bosonic & fermionic \\ \hline
gravitational & $1$ & $g_{\mu\nu}$ & $\psi_\mu$ \\ \hline
vector & $\mathrm{dim}(G\times E_8)$ & $A^A_\mu$ & $\chi^A$ \\ \hline
\multirow{2}{*}{chiral moduli} & $h^{(1,1)}$ & $t^i$ & $\xi^i$ \\ \cline{2-4}
& $h^{(1,2)}$ & $z^a$ & $\zeta^a$ \\ \hline
linear & $1$ & $B_2,\phi$ & $\lambda$ \\ \hline
\multirow{2}{*}{chiral matter} & $h^{(1,1)}$ & $A^i$ & $\chi^i$ \\ \cline{2-4}
& $h^{(1,2)}$ & $A^a$ & $\chi^a$ \\
\hline
\end{tabular}
\end{center}
\caption{\label{tab-4d-spec}$N=1$ spectrum in $D=4$.}
\end{table}

\subsection{Fermionic spectrum}\label{sec-spec-fer}

In order to derive the fermionic spectrum we first need to discuss the decomposition of a ten-dimensional Majorana-Weyl spinor $\hat\epsilon$ in the background (\ref{backg}). One has
\begin{equation}
\hat\epsilon = \bar\epsilon \otimes \eta_+ + (\bar\epsilon)^\ast \otimes (\eta_+)^\ast = \bar\epsilon \otimes \eta_+ + \epsilon \otimes \eta_-\ ,
\label{spindecomp}
\end{equation}
where $\epsilon$ and $\bar\epsilon$ are Weyl spinors of $M_4$ while $\eta_\pm$ are Weyl spinors of $\hat Y$. The $^\ast$ stands for complex conjugation and we summarize our spinor conventions in appendix \ref{app-conv}.

On generic manifolds $\hat Y$ with structure group $\mathrm{SO}(6)$ $\eta$ transforms in the $\mathbf{4}$ of $\mathrm{SO}(6)$. However, on manifolds with $\mathrm{SU}(3)$ structure the $\mathbf{4}$ decomposes as $\mathbf{4} \to \mathbf{3} \oplus \mathbf{1}$. In the following, $\eta$ is the normalized ($\eta_\pm^\dag \eta_\pm = 1$) singlet in this decomposition.

The next step is to decompose the ten-dimensional fermions, the gravitino $\hat\psi_M$, the gauginos $\hat\chi$ and the dilatino $\hat\lambda$ in the background (\ref{backg}) using (\ref{spindecomp}). For the gaugino $\hat\chi$, transforming in the adjoint of $G\times E_8$, and $\hat\lambda$, this is straightforward and reads
\begin{equation}
\begin{aligned}
\hat\chi &= \chi\otimes\eta_- +  \bar\chi\otimes\eta_+\ ,\\
\hat\lambda &= \lambda\otimes\eta_+ + \bar\lambda\otimes\eta_-\ ,
\end{aligned}
\label{chilambdadecomp}
\end{equation}
where $\chi$ is the Weyl spinor corresponding to the four-dimensional gaugino while $\lambda$ denotes the Weyl spinor corresponding to the four-dimensional dilatino.

Analogously to (\ref{adecomp}), we expand the chiral matter fields in terms of forms which transform in the $\mathbf{6}$ or in the $\mathbf{8}\oplus\mathbf{1}$ of $\mathrm{SU}(3)$
\begin{equation}
\hat\chi_\alpha = \chi^i \otimes (\omega_i)_{\alpha\bar\beta}
\gamma^{\bar\beta} \eta_+ + \frac1{\Vert\Omega\Vert^2}\, \bar\chi^a
\otimes (\bar\rho_a)_{\alpha\bar\beta\bar\gamma}
{\Omega_\delta}^{\bar\beta\bar\gamma} \gamma^\delta \eta_-\ .
\label{chidecomp}
\end{equation}
Note that $\chi$ is the four-dimensional gaugino transforming in the adjoint representation of the gauge group $G\times E_8$ while $\chi^i$ and $\chi^a$ are four-dimensional chiral matter fermions.

Finally, we see from table \ref{tab-su3-fer} that in the decomposition of the ten-dimensional gravitino $\hat\psi_M$ we only want to keep the singlet in $\hat\psi_\mu$ and the $\mathbf{6}$ and $\mathbf{8}\oplus\mathbf{1}$ representations of $\mathrm{SU}(3)$ in $\hat\psi_m$. Thus we decompose
\begin{equation}
\begin{aligned}
\hat\psi_\mu &= \psi_\mu\otimes\eta_- + \bar\psi_\mu\otimes\eta_+\ ,\\
\hat\psi_\alpha &= \xi^i \otimes (\omega_i)_{\alpha\bar\beta} \gamma^{\bar\beta} \eta_+ + \frac1{\Vert\Omega\Vert^2}\, \bar\zeta^a \otimes (\bar\rho_a)_{\alpha\bar\beta\bar\gamma} {\Omega_\delta}^{\bar\beta\bar\gamma} \gamma^\delta \eta_-\ ,
\end{aligned}
\label{psidecomp}
\end{equation}
where $\xi^i$ ($\zeta^a$) are $h^{(1,1)}$ ($h^{(1,2)}$) four-dimensional gauge neutral Weyl fermions. Together with the bosonic fields of the previous section, the fermions combine into supermultiplets as summarized in table \ref{tab-4d-spec}.

\section{The low energy effective action}\label{sec-eff-act}

In this section we compute the four-dimensional low energy effective action by a Kaluza-Klein reduction. We start from the ten-dimensional $N=1$ supergravity action of the heterotic string \cite{Romans:1985xd}. It encodes the low energy dynamics of the gravitational and vector multiplets given in table \ref{tab-10d-spec}. The action can be split into three distinct contributions, $S^{(10)} = S_\mathrm{b} + S_\mathrm{f} + S_\mathrm{int}$. $S_\mathrm{b}$ includes the purely bosonic terms and reads\footnote{Throughout this paper we will use the following shorthands, $\int_{10} \equiv \int d^{10}x \sqrt{-\hat G_{10}}$, $\int_4 \equiv \int d^{4}x \sqrt{-g_{4}}$ and $\int_6 \equiv \int d^{6}y \sqrt{g_{6}}$.}
\begin{equation}
S_\mathrm{b} = -\int_{10} \Bigl[ \tfrac12 \hat R + \tfrac1{24} e^{-\hat\phi} \hat H_{MNP} \hat H^{MNP} + \tfrac12 \partial_M \hat\phi \partial^M \hat\phi + \tfrac14 e^{-\frac{\hat\phi}2} \hat F^{\hat A}_{MN} \hat F^{\hat A,MN} \Bigl]\ .
\label{sb}
\end{equation}
On the other hand, $S_\mathrm{f}$ contains the kinetic terms for the fermions
\begin{equation}
S_\mathrm{f} = - \int_{10} \Bigl[ \hat{\bar\psi}_M \Gamma^{MNP} D_N \hat\psi_P + \hat{\bar\lambda} \Gamma^M D_M \hat\lambda + \hat{\bar\chi}^{\hat A} \Gamma^M D_M \hat\chi^{\hat A} \Bigl]\ ,
\label{sf}
\end{equation}
while $S_\mathrm{int}$ contains the interactions
\begin{equation}
\begin{aligned}
S_\mathrm{int} = - \int_{10} & \Bigl[ \tfrac1{\sqrt2}\, \partial_N \hat\phi (\hat{\bar\psi}_M \Gamma^N \Gamma^M \hat\lambda) - \tfrac12 e^{-\frac{\hat\phi}4} \hat F^{\hat A}_{MN} \hat{\bar\chi}^{\hat A} \Gamma^Q \Gamma^{MN} (\hat\psi_Q + \tfrac{\sqrt2}{12}\Gamma_Q\hat\lambda) \\
&{} + \tfrac1{24} e^{-\frac{\hat\phi}2} \hat H_{MNP} (\hat{\bar\psi}_Q \Gamma^{QMNPR} \hat\psi_R + 6 \hat{\bar\psi}^M \Gamma^N \hat\psi^P - \sqrt2\hat{\bar\psi}_Q \Gamma^{MNP} \Gamma^Q \hat\lambda) \\
&{} + \tfrac1{24} e^{-\frac{\hat\phi}2} \hat H_{MNP} \hat{\bar\chi}^{\hat A} \Gamma^{MNP} \hat\chi^{\hat A} + \textrm{four Fermi terms} \Bigl]\ .
\end{aligned}
\label{sint}
\end{equation}

We have given $S^{(10)}$ in the Einstein frame and $\hat R$ is the Ricci scalar in that frame. $\Gamma^{M_1\ldots M_p} = \frac1{p!} \Gamma^{[M_1}\ldots \Gamma^{M_p]}$ denote totally antisymmetrized products of $\Gamma$-matrices. $\hat F^A_{MN}$ is the field strength for the gauge boson $\hat A^A_M$ and $\hat H$ is the modified three-form field strength of $\hat B$ defined as
\begin{equation}
\hat H_3 = d\hat B_2 - \omega_3^{\mathrm{YM}} + \omega_3^{\mathrm{L}}\ ,
\end{equation}
where $\omega_3^{\mathrm{YM}}$ is the Yang-Mills Chern-Simon three-form and $\omega_3^{\mathrm{L}}$ is the Lorentz Chern-Simon three-form.

\subsection{The kinetic terms in the $D=4$ bosonic action}\label{bosonicK}

Let us first compute the kinetic terms of the bosons in the $D=4$ effective action. For Calabi-Yau compactifications this is a well-known result and has been computed for example in references \cite{Strominger:1985ks,Dixon:1989fj,Candelas:1990pi}. For manifolds with $\mathrm{SU}(3)$ structure the analogous computation has been performed in \cite{Gurrieri:2002wz,Grana:2005ny,Micu:2004tz,Gurrieri:2004dt,Gurrieri:2007jg}. Therefore we can be brief and basically just recall the results.

Since the $D=4$ effective action has $N=1$ supersymmetry it can be expressed in terms of a real K\"ahler potential $K$, the holomorphic gauge kinetic function $f$ and the holomorphic superpotential $W$ \cite{Wess:1992cp}. (Some of the couplings relevant to this paper are recorded in appendix \ref{app-sugra}.)

One inserts the expansions (\ref{b2}) -- (\ref{adecomp}) into (\ref{sb}), performs a Weyl rescaling of the four-dimensional metric and defines the four-dimensional dilaton $\phi^{(4)}$ according to
\begin{equation}
g_{\mu\nu} \to e^{\frac32\phi} \mathcal{K}^{-1} g_{\mu\nu}\ , \qquad \phi^{(4)} = \phi - \tfrac12 \ln \mathcal{K}\ ,
\label{weylresc}
\end{equation}
where $\mathcal{K}$ is the volume of $\hat Y$ given by
\begin{equation}
\mathcal{K} = \tfrac16 \int_{\hat Y} J\wedge J\wedge J\ , \qquad J = v^i \omega_i\ .
\label{volcy}
\end{equation}
With these redefinitions one obtains an action of the form (\ref{bosonicaction}) with a gauge kinetic function
\begin{equation}
f = S\ , \qquad S = \tfrac12 e^{-2\phi^{(4)}} + \tfrac\im 2\, a\ ,
\label{gaugekineticf}
\end{equation}
where $a$ is the dual of $B_2$. Furthermore the K\"ahler metric $g_{I\bar J}$ on the field space for the moduli multiplets is block-diagonal with the non-trivial entries
\begin{equation}
g_{S\bar S} = \frac1{(S + \bar S)^2}\ , \qquad g_{ij} = \frac1{4\mathcal{K}} \int_{\hat Y} \omega_i \wedge \ast \omega_j\ , \qquad g_{a\bar b} = \frac{\int_{\hat Y} \rho_a \wedge \bar\rho_b}{\int_{\hat Y} \Omega \wedge \bar \Omega}\ ,
\label{metrics}
\end{equation}
where $g_{S\bar S}$ is the metric for the dilaton, $g_{ij}$ is the metric for the $h^{(1,1)}$ chiral multiplets $(t^i, \xi^i)$ and $g_{a\bar b}$ is the metric for the $h^{(1,2)}$ chiral multiplets $(z^a, \zeta^a)$. Each metric can be shown to be a K\"ahler metric so that the K\"ahler potential is the sum of three terms
\begin{equation}
K=K_S + K_J + K_\Omega\ ,
\label{k}
\end{equation}
where
\begin{equation}
\begin{aligned}
e^{-K_S} = (S + \bar S)^2\ , \qquad
e^{-K_J} = \mathcal{K}\ , \qquad
e^{-K_\Omega} = \im \int_{\hat Y} \Omega \wedge \bar \Omega\ .
\end{aligned}
\label{keach}
\end{equation}
In fact both $K_J$ and $K_\Omega$ define special K\"ahler manifolds in that they can be derived from holomorphic prepotentials. Further details can be found in appendix \ref{app-geom}.

For the matter scalars $A^i, A^a$ one computes the moduli-dependent metrics $Z_{ij}$ and $Z_{a\bar b}$. From the Kaluza-Klein reduction one finds straightforwardly by inserting (\ref{adecomp}) into the last term in (\ref{sb})
\begin{equation}
Z_{ij} = 4e^{-\frac\phi2} {\tilde g}_{ij}\ , \qquad Z_{a\bar b} = 4e^{-\frac\phi2}g_{a\bar b}\ .
\end{equation}
However a holomorphic superpotential requires a further rescaling \cite{Gurrieri:2007jg}
\begin{equation}
A^i \to \tfrac12\Vert\Omega\Vert^{-\frac13} A^i = \tfrac12 e^{-\frac\phi4} e^{\tfrac16(K_\Omega - K_J)} A^i\ , \qquad A^a \to \tfrac12\Vert\Omega\Vert^{\frac13} A^a = \tfrac12 e^{\frac\phi4} e^{\tfrac16(K_J - K_\Omega)} A^a\ ,
\label{arescal}
\end{equation}
which results in the metric, derived in \cite{Dixon:1989fj} via conformal field theory,
\begin{equation}
Z_{ij} = e^{\tfrac13(K_\Omega - K_J)} g_{ij}\ , \qquad Z_{a\bar b} = e^{\tfrac13(K_J - K_\Omega)} g_{a\bar b}\ .
\label{Z}
\end{equation}
Shortly we will see that it is precisely this normalization of matter
fields which leads to appropriate expressions for the Yukawa
couplings.
Note that \eqref{Z} only gives the moduli dependent part of the metric
which does not include the dependence on $A^i$ and $A^a$ itself. In
particular we are not computing terms of the form $h_{ia}(t,\bar t, z,\bar
z) A^i A^a$ in the K\"ahler potential. These terms have been
determined in \cite{Gurrieri:2007jg}.

\subsection{The kinetic terms in the $D=4$ fermionic action}\label{fermionicK}

Let us now turn to the computation of the kinetic terms of the fermions. Due to the supersymmetry this will not add any new information but will merely be a consistency check on the method and the couplings computed in the previous section. However, since we intend to compute the superpotential and $D$-terms via fermionic couplings it is mandatory to properly fix the normalization of the fermionic terms. Surprisingly, to our knowledge this has not appeared in the literature so far, not even for Calabi-Yau compactifications.

Since in this first step we are going to compute the kinetic terms we only need to focus on those terms in (\ref{sf}) which contain a space-time derivative $D_\mu$. The $\Gamma$-matrices are decomposed as
in (\ref{gammadecomp}). (Our spinor conventions are summarized in appendix \ref{app-conv}.) As a consequence of (\ref{annihil}), terms like $\eta_\pm^\dag \gamma^\alpha\ldots \gamma^{\bar\beta} \eta_\pm$ vanish unless they have an equal number of holomorphic and antiholomorphic $\gamma$-matrices. These terms in turn can be simplified by using (\ref{cliffholo}).
\begin{equation}
\eta_+^\dag \gamma^\gamma \gamma^{\bar\alpha\beta} \gamma^{\bar\delta}
\eta_+ = 4 g^{\gamma\bar\alpha} g^{\beta\bar\delta} - 2
g^{\beta\bar\alpha} g^{\gamma\bar\delta}\ .
\label{wick}
\end{equation}
The kinetic terms of $\psi_\mu$, $\lambda$ and $\chi$ follow straightforwardly by inserting (\ref{chilambdadecomp}), (\ref{chidecomp}) and (\ref{psidecomp}) into (\ref{sf}). The only complication arises from terms involving $\hat\Psi_\alpha$. In their reduction one encounters the integrals
\begin{equation}
\begin{aligned}
\int_6 (\omega_i)_{\alpha\bar\beta} (\omega_j)_{\gamma\bar\delta} \big[ g^{\alpha\bar\delta} g^{\gamma\bar\beta} - 2 g^{\alpha\bar\beta} g^{\gamma\bar\delta} \big] &= 2 \tilde{\mathcal{K}}_{ij} + 4 \tilde{\mathcal{K}} \tilde{g}_{ij}\ , \\ 
\frac1{\Vert\Omega\Vert^2} \int_6 (\rho_a)_{\bar\alpha\beta\gamma} (\bar\rho_b)_{\delta\bar\epsilon\bar\zeta}
\bar\Omega_{\bar\sigma}^{\beta\gamma} \Omega_{\tau}^{\bar\epsilon\bar\zeta} \big[ g^{\delta\bar\alpha}
g^{\tau\bar\sigma} - 2 g^{\delta\bar\sigma} g^{\tau\bar\alpha} \big] &= - \frac{4i}{\Vert\Omega\Vert^2} \int_{\hat Y} \rho_a \wedge \bar\rho_b = -4 \tilde{\mathcal{K}} g_{a\bar b}\ ,
\label{int6}
\end{aligned}
\end{equation}
where we abbreviated $\tilde{\mathcal{K}}_{ij} = \int_{\hat Y} \omega_i \wedge \omega_j \wedge \tilde J$ and used (\ref{metrics}) in the Einstein frame, \eqref{volcy} and (\ref{wick}). (Quantities with a
tilde depend on the moduli $\tilde v$ defined in \eqref{deltag} and \eqref{frames} or in other words they depend on the deformations of the metric in the string frame.) With these formulas at hand one arrives at
\begin{equation}
\begin{aligned}
S_\mathrm{f} = \int_4 & \Bigl[ 2\tilde{\mathcal{K}}\epsilon^{\mu\rho\nu\lambda} \bar\psi_\mu \bar\sigma_\lambda D_\rho \psi_\nu + \im\tilde{\mathcal{K}}_i \xi^i \sigma^{[\mu} \bar\sigma^{\nu]} D_\mu\psi_\nu \\
&+ \im\tilde{\mathcal{K}}_i \bar\psi_\mu \bar\sigma^{[\mu} \sigma^{\nu]} D_\nu\bar\xi^i + 8 \im \{ \tilde{\mathcal{K}}_{ij} + 2 \tilde{\mathcal{K}} \tilde{g}_{ij} \} \bar\xi^i \bar\sigma^\mu D_\mu \xi^j \\
& {} - 16\im \tilde{\mathcal{K}} g_{\bar ab} \bar\zeta^a \bar\sigma^\mu D_\mu \zeta^b - 2\im \tilde{\mathcal{K}} \bar\lambda \bar\sigma^\mu D_\mu \lambda - 2 \im \tilde{\mathcal{K}} \bar\chi \bar\sigma^\mu D_\mu \chi \\
& {} - 16 \im \tilde{\mathcal{K}} {\tilde g}_{ij} \bar\chi^i
\bar\sigma^\mu D_\mu \chi^j - 16 \im \tilde{\mathcal{K}} g_{\bar ab}
\bar\chi^a \bar\sigma^\mu D_\mu \chi^b \Bigl]\ ,
\end{aligned}
\label{sreduc}
\end{equation}
where $\tilde{\mathcal{K}}_i = \int_{\hat Y} \omega_i \wedge \tilde J \wedge \tilde J$.

The next step is to perform the Weyl rescaling of the metric as in (\ref{weylresc}). Since the $\sigma^\mu$ are defined with a vierbein they also rescale. In addition,  all fermionic fields Weyl rescale as follows \cite{Wess:1992cp}
\begin{equation}
\begin{aligned}
\sigma^\mu \to \tilde{\mathcal{K}}^{\tfrac12} \sigma^\mu\ , \qquad
\psi^\mu \to \tilde{\mathcal{K}}^{-\tfrac14} \psi_\mu\ , \qquad \xi^i
\to \tilde{\mathcal{K}}^{\tfrac14} \xi^i\ , \qquad \zeta^a \to \tilde{\mathcal{K}}^{\tfrac14} \zeta^a\ , \\
\lambda \to \tilde{\mathcal{K}}^{\tfrac14} \lambda\ , \qquad \chi \to \tilde{\mathcal{K}}^{\tfrac14} \chi\ , \qquad \chi^i \to \tilde{\mathcal{K}}^{\tfrac14} \chi^i\ , \qquad \chi^a \to \tilde{\mathcal{K}}^{\tfrac14} \chi^a\ .
\label{resc1}
\end{aligned}
\end{equation}
Inspecting the Lagrangian in (\ref{sreduc}) we also see that the kinetic terms are not yet diagonal. They can be diagonalized by shifting the gravitino as follows 
\begin{equation}
\psi_\mu \to \psi_\mu + \sigma_\mu\ \frac{\tilde{\mathcal{K}}_i \bar\xi^i}{2\tilde{\mathcal{K}}}\ .
\label{gravshift}
\end{equation}
Inserting (\ref{resc1}) and (\ref{gravshift}) into (\ref{sreduc}) we arrive at
\begin{equation}
\begin{aligned}
S_\mathrm{f} = -\im \int_4 & \Bigl[ 2 \im \epsilon^{\mu\rho\nu\lambda}  \bar\psi_\mu \bar\sigma_\lambda D_\rho \psi_\nu - 8 \bar\xi^i \bar\sigma^\mu D_\mu \xi^j \Big( \frac{\tilde{\mathcal{K}}_{ij}} {\tilde{\mathcal{K}}} - \frac{3\tilde{\mathcal{K}}_i \tilde{\mathcal{K}}_j} {8\tilde{\mathcal{K}}^2} + 2 \tilde{g}_{ij} \Big) \\
&{} + 16 \bar\zeta^a \bar\sigma^\mu D_\mu \zeta^b \tilde{\mathcal{K}} g_{\bar ab} + 2 \tilde{\mathcal{K}} \bar\lambda \bar\sigma^\mu D_\mu \lambda + 2 \bar\chi \bar\sigma^\mu D_\mu \chi \\
& {} + 16 {\tilde g}_{ij} \bar\chi^i \bar\sigma^\mu D_\mu \chi^j + 16
g_{\bar ab} \bar\chi^a \bar\sigma^\mu D_\mu \chi^b \Bigl]\ .
\end{aligned}
\label{sreduc2}
\end{equation}

To bring the kinetic terms of all fermions to the standard form (\ref{fermionicaction}) as dictated by $N=1$ supergravity the fermionic fields need to be further rescaled as follows,
\begin{equation}
\begin{aligned}
\psi_\mu \to \tfrac1{\sqrt2} \psi_\mu\ , \qquad \xi^i \to \tfrac14 e^{-\frac\phi2} \Bigl( \xi^i - \frac{\tilde{v}^i\, \tilde{\mathcal{K}}_j   \xi^j}{12\tilde{\mathcal{K}}}  \Bigl)\ , \qquad \zeta^a \to \tfrac14 \zeta^a\ , \qquad \lambda \to \tfrac{1}{\sqrt2}\, e^{2\phi^{(4)}}\,\lambda\ , \\
\chi \to \tfrac1{\sqrt2}\, e^{-\phi^{(4)}} \chi\ , \qquad \chi^i \to \tfrac14 \,e^{-\frac{\phi}2} e^{\frac16 (K_{\Omega} - K_J)} \chi^i\ , \qquad \chi^a \to \tfrac14\, e^{\frac16 (K_J - K_{\Omega})} \chi^a\ .
\end{aligned}
\label{resc2}
\end{equation}
Notice that using $\tilde{\mathcal{K}}_i \tilde v^i = 6\tilde{\mathcal{K}}$ the combination $\tilde{\mathcal{K}}_i \xi^i$ transforms in the much simpler way
\begin{equation}
\tilde{\mathcal{K}}_i \xi^i \to \tfrac18\, e^{-\frac\phi2} \tilde{\mathcal{K}}_i \xi^i\ .
\end{equation}
It is straightforward to check that after substitution of (\ref{resc2}) into (\ref{sreduc2}) we arrive at
\begin{equation}
\begin{aligned}
S_\mathrm{f} = - \im \int_4 & \Bigl[ \im \epsilon^{\mu\rho\nu\lambda}  \bar\psi_\mu \bar\sigma_\lambda D_\rho \psi_\nu + g_{ij} \bar\xi^i \bar\sigma^\mu D_\mu \xi^j + g_{\bar ab} \bar\zeta^a \bar\sigma^\mu D_\mu \zeta^b \\ 
&+ g_{S\bar S} \bar\lambda \bar\sigma^\mu D_\mu \lambda + \mathrm{Re}f \bar\chi\sigma^\mu D_\mu \chi + Z_{ij} \bar\chi^i \bar\sigma^\mu D_\mu\chi^j + Z_{\bar ab} \bar\chi^a \bar\sigma^\mu D_\mu \chi^b \Bigl]\ ,
\end{aligned}
\label{sreduc3}
\end{equation}
in accord with the standard form of $N=1$ supergravity.\footnote{Note that in the derivation of (\ref{sreduc3}) we have ignored all terms where space-time derivatives act on bosonic terms. They should combine into appropriate covariant derivatives as given in \cite{Wess:1992cp} which, however, we did not explicitly check.}

\subsection{Yukawa Couplings}\label{Yukawa}

After having determined the kinetic terms let us now turn to the Yukawa couplings. They arise from the kinetic terms of the ten-dimensional gaugino $\hat\chi^A$ given as the last term in (\ref{sf}). More specifically they arise from terms inside the covariant derivative which have the structure
\begin{equation}
S_{\mathrm{Yukawa}} = \int_{10} {f^A}_{BC}{\hat{\bar\chi}}^A \Gamma^M {\hat A}^B_M {\hat\chi}^C.
\end{equation}
Restricting to the matter fields and inserting the expansions (\ref{adecomp}) and (\ref{chidecomp}) one arrives at
\begin{equation}
\begin{aligned}
S_{\mathrm{Yukawa}} = \int_4  \Big[ {\chi}^a \cdot \chi^c \cdot A^b \frac{4}{\Vert\Omega\Vert^4} \int_6 {\Omega_\epsilon}^{\bar\gamma\bar\alpha} (\bar\rho_a)_{\beta\bar\gamma\bar\alpha} {(\bar\rho_b)_\alpha}^{\delta\beta} {(\bar\rho_c)_\delta}^{\epsilon\alpha} + \\
{} + {\chi}^i\cdot \chi^k \cdot A^j \int_{\hat Y} \omega_i \wedge \omega_j \wedge \omega_k + \textrm{h.c.} \Big]\,
\label{yukawa}
\end{aligned}
\end{equation}
where we used (\ref{omegaeps}). The dots indicate that the four-dimensional fields are contracted with an invariant tensor of the group $G$ which arise from the decomposition as discussed in section \ref{sec-4d-spec} of the ten-dimensional $E_8$ structure constants ${f^A}_{BC}$.

The first term in (\ref{yukawa}) can be rewritten as 
\begin{equation}
\begin{aligned}
\frac1{2\Vert\Omega\Vert^4} \int_6 {\Omega_\epsilon}^{\bar\gamma\bar\alpha}  (\bar\rho_a)_{\beta\bar\gamma\bar\alpha} {(\bar\rho_b)_\alpha}^{\delta\beta} {(\bar\rho_c)_\delta}^{\epsilon\alpha} & = \int_{\hat Y} \Omega \wedge \rho_a^\alpha \wedge \rho_b^\beta \wedge \rho_c^\gamma \Omega_{\alpha\beta\gamma} \\
& = \frac{\partial^3\mathcal{G}}{\partial z^a\partial z^b\partial z^c},
\label{g}
\end{aligned}
\end{equation}
where $\rho_a^\alpha$ is defined as follows
\begin{equation}
\rho_a^\alpha = \frac1{2\Vert\Omega\Vert^2} \bar\Omega^{\alpha\beta\gamma}
(\rho_a)_{\beta\gamma\bar\alpha}dz^{\bar\alpha}.
\end{equation}
The second equation in (\ref{g}) uses (\ref{omegadecomp}) and has been proved in \cite{Candelas:1987se,Candelas:1990pi} for Calabi-Yau manifolds. We expect it to hold also for manifolds with $SU(3)$ structure. 

Finally by using (\ref{arescal}), (\ref{resc2}) and (\ref{cubic}) one arrives at
\begin{equation}
S_{\mathrm{Yukawa}} = \int_4  e^{\frac{K}{2}} \Big[ \chi^a \cdot \chi^c \cdot A^b \frac{\partial^3 \mathcal{G}}{\partial z^a\partial z^b \partial z^c} + \chi^i \cdot \chi^k \cdot A^j \frac{\partial^3 F}{\partial t^i \partial t^j \partial t^k} +\textrm{h.c.}\Big].
\end{equation}
Using \eqref{appsint} this corresponds to a contribution in the superpotential which is of the form
\begin{equation}
W= \frac 13\frac{\partial^3 \mathcal{G}}{\partial z^a\partial z^b \partial
z^c}\, A^a A^bA^c+\frac 13
\frac{\partial^3 F}{\partial t^i \partial t^j \partial t^k}\,
A^iA^jA^k \ .
\end{equation}
The requirement that this term is holomorphic fixes the rescalings \eqref{arescal}. In \cite{Gurrieri:2007jg,Correia:2007sv} it was shown that this term can also be expressed as an appropriate integral on the internal manifold.

\subsection{Gravitino mass term and $F$-terms}\label{Fterms}

So far we computed the kinetic terms and the Yukawa couplings in the four-dimensional effective theory by a Kaluza-Klein reduction. Now we continue in this spirit and derive the gravitino mass term and the $F$-terms related to non-trivial fluxes and/or torsion. In principle there are two ways to do this computation. One can reduce the bosonic part of the ten-dimensional action (\ref{sb}) -- (\ref{sint}) and derive the scalar potential $V$. From the supergravity relation (\ref{v}) one can then infer the superpotential and the $D$-terms. However, this procedure is problematic since $W$ and its derivatives enter quadratically in $V$ and thus cannot be computed reliably within the approximation used. However in the fermionic couplings both $W$ and its derivatives appear linearly and therefore can be obtained more easily \cite{Gurrieri:2004dt,Jockers:2005zy}.

The $N=1$ supergravity Lagrangian given in (\ref{appsint}) tells us that $W$ can be computed from the gravitino mass term while the derivatives of $W$ (the $F$-terms) can be computed from the couplings of the gravitino to the chiral fermions. In the following we are going to determine these couplings via Kaluza-Klein reduction.

The contributions to the gravitino mass term and to the $F$-terms arise from two different sources. On the one hand they come from the reduction of $S_{\mathrm{f}}$ given in (\ref{sf}) when no space-time derivative $D_\mu$ is present and the internal derivative $D_m$ acts on the spinor $\eta$ in the expansion of the fermions (\ref{chilambdadecomp}) -- (\ref{psidecomp}). In this case, they will be proportional to certain torsion components of the $\mathrm{SU}(3)$ manifold. The second possibility is that such terms arise from the reduction of $S_\mathrm{int}$ given in (\ref{sint}) when the three-form field strength $H_3$ takes a non-trivial background value (background flux) on the manifold $\hat Y$. Let us discuss both cases starting with the contribution arising from $H_3$-flux.

\subsubsection*{Contribution from $H_3$-flux}

In this case the gravitino mass term descends from the first term in the second line of (\ref{sint}) where both gravitini carry indices $\mu, \nu = 0, \ldots, 3$. Inserting (\ref{psidecomp}) and using (\ref{jomega}) one finds
\begin{equation}
\begin{aligned}
S_{\mathrm{int},H} &= - \tfrac1{24} \int_{10} e^{-\frac{\hat \phi}2} \hat H_{MNP} \hat{\bar{\psi}}_L \Gamma^{LMNPQ} \hat\psi_Q \\
&= \tfrac1{48} \int_4 \bar\psi_\mu \bar\sigma^{[\mu}\sigma^{\nu]} \bar\psi_\nu e^{-\frac\phi2}\int_6 H_{\bar\alpha\bar\beta\bar\gamma} \eta_-^\dag \gamma^{\bar\alpha\bar\beta\bar\gamma} \eta_+ + \mathrm{h.c.} \\
&= - \tfrac18 \int_4 \bar\psi_\mu \bar\sigma^{[\mu}\sigma^{\nu]} \bar\psi_\nu e^{-\frac\phi2}\int_{\hat Y} \Omega_\eta \wedge H + \mathrm{h.c.}\ ,
\end{aligned}
\label{sh}
\end{equation}
where we only display the contribution to the gravitino mass. Performing the Weyl rescaling using (\ref{weylresc}), (\ref{resc1}) and (\ref{resc2}), and using (\ref{absomega}) we obtain
\begin{equation}
S_{\mathrm{int},H} = - \tfrac14 \int_4 \bar\psi_\mu \bar\sigma^{\mu\nu} \bar\psi_\nu e^{\frac K2}\int_{\hat Y} \Omega \wedge H + \mathrm{h.c.}\ ,
\label{sgravmassh}
\end{equation}
where $\bar\sigma^{\mu\nu} = \frac14 \bar\sigma^{[\mu} \sigma^{\nu]}$. Comparing with (\ref{appsint}) we see that the contribution to the superpotential arising from the background $H_3$-flux is given by
\begin{equation}
W_H = \tfrac14 \int_{\hat Y} \Omega \wedge H \ ,
\label{wh}
\end{equation}
a result computed previously in references \cite{Behrndt:2000zh,Becker:2002jj,Cardoso:2003af}. Note that this derivation provides an independent check on the K\"ahler potentials (\ref{k}), (\ref{keach}) which we explicitly used in (\ref{sh}).

Let us now proceed to the computation of the derivatives of $W$ or in other words the $F$-terms. They arise from the same ten-dimensional term $S_{\mathrm{int},H}$ in (\ref{sh}) but this time choosing one of the ten-dimensional gravitini to carry an internal index. There is also an additional contribution coming from the insertion of the gravitino shifting (\ref{gravshift}) in (\ref{sgravmassh}). Inserting the decomposition (\ref{psidecomp}) one finds
\begin{equation}
\begin{aligned}
S_{\mathrm{int},H} =\ & - \tfrac1{48} \int_4\, \frac{\tilde{\mathcal{K}}_i}{2\tilde{\mathcal{K}}}\, \xi^i \sigma^\mu \bar\psi_\mu \int_6 H_{\bar\alpha\bar\beta\bar\gamma} \eta_-^\dag \gamma^{\bar\alpha\bar\beta\bar\gamma} \eta_+  \\
& {} + \tfrac14 \int_4 \zeta^a \sigma^\mu \bar\psi_\mu e^{-\frac\phi2}
\frac1{\Vert\Omega\Vert} \int_6 (\rho_a)_{\bar\alpha\beta\gamma}
H_{\delta\bar\epsilon\bar\zeta} (\bar\Omega_\eta)^{\delta\beta\gamma}
(\Omega_\eta)^{\bar\alpha\bar\epsilon\bar\zeta} + \mathrm{h.c.}\ ,
\end{aligned}
\end{equation}
where this time only the contributions relevant to the $F$-terms are shown. In computing this expression the following property, derived in \cite{KashaniPoor:2006si}, was used
\begin{equation}
(\omega_i)_{\alpha\bar\beta} g^{\alpha\bar\beta} = \frac{\im\tilde{\mathcal K}_i}{2\tilde{\mathcal K}}\ ,
\label{omegagcst}
\end{equation}
which is therefore independent of the internal coordinates. Using (\ref{jomega}), (\ref{absomega}), (\ref{astomega}) and the Weyl rescaling according to (\ref{weylresc}), (\ref{resc1}) and (\ref{resc2}) one obtains 
\begin{equation}
S_H = - \tfrac\im{4\sqrt2} \int_4 e^{\frac K2}\Bigl[ \xi^i \sigma^\mu \bar\psi_\mu\, \frac{\im\mathcal{K}_i}{4\mathcal{K}} \int_{\hat Y} \Omega\wedge H + \zeta^a \sigma^\mu \bar\psi_\mu \int_{\hat Y} \rho_a \wedge H \Bigl] + \mathrm{h.c.}\ .
\label{sh1}
\end{equation}
It is straightforward to check from (\ref{kahlerderiv}), (\ref{keach}), (\ref{volcy}) and (\ref{deromega}) that in the absence of torsion (or in other words for $d\omega_i = 0$) the K\"ahler derivatives of $W_H$ as obtained in (\ref{wh}) are given by
\begin{equation}
D_i W_H = \frac{\im\mathcal{K}_i}{4 \mathcal{K}}\ W_H\ , \qquad D_a W_H = \tfrac14 \int_{\hat Y} \rho_a \wedge H\ .
\label{dwh}
\end{equation}
We therefore conclude 
\begin{equation}
S_{\mathrm{int},H} = - \tfrac\im{\sqrt2} \int_4 e^{\frac K2} \Bigl\{ \xi^i \sigma^\mu \bar\psi_\mu D_i W_H + \zeta^a \sigma^\mu \bar\psi_\mu D_a W_H \Bigl \} + \mathrm{h.c.}\ ,
\end{equation}
which, after comparing with (\ref{appsint}), shows the consistency with $N=1$ supergravity.

Finally, there also is a gravitino-dilatino coupling which is obtained from the appropriate term in (\ref{sint})
\begin{equation}
\begin{aligned}
S_{\mathrm{int},\psi\lambda} &= \tfrac{\sqrt2}{24} \int_{10} e^{-\frac{\hat\phi}2} \hat H_{mnp} \hat{\bar \psi}_\mu \Gamma^{mnp} \Gamma^\mu \hat\lambda \\
&= \tfrac {\im\sqrt2}4 \int_4 \lambda \sigma^\mu \bar\psi_\mu e^{-\frac\phi2} \int_{\hat Y} \Omega_\eta \wedge H + \mathrm{h.c.}\ ,
\end{aligned}
\end{equation}
where we inserted (\ref{chilambdadecomp}) and (\ref{psidecomp}). Following a similar procedure as for the other $F$-terms and performing the Weyl rescaling one gets
\begin{equation}
S_{\mathrm{int},\psi\lambda} = - \tfrac\im{\sqrt2} \int_4 \lambda \sigma^\mu \bar\psi_\mu e^{\frac K2} D_S W_H \ ,
\label{shlambda}
\end{equation}
where we used
\begin{equation}
D_S W_H = - e^{2\phi^{(4)}} W_H\ ,
\end{equation}
as can be derived from (\ref{kahlerderiv}), (\ref{gaugekineticf}), (\ref{metrics}) and (\ref{keach}). This result is again in agreement with supergravity if we compare with (\ref{appsint}).

\subsubsection*{Contribution from torsion}

In addition to $H_3$-flux also the torsion of the manifold $\hat Y$ appears in $W$. These contributions arise from (\ref{sf}) precisely when an internal derivative acts on the internal spinor $\eta$. Before we do the reduction let us briefly recall the structure of these derivatives as determined in \cite{Grana:2004bg}. One decomposes $D_m \eta_\pm$ into a basis $(\eta, \gamma^7 \eta, \gamma^n \eta)$ and defines the tensors $q_m$, $q_m^\prime$ and $q_{mn}$ via
\begin{equation}
\begin{aligned}
&D_m \eta_+ = q_m \eta_+ + \im q^\prime_m \gamma^7 \eta_+ + \im q_{mn} \gamma^n \eta_-\ , \\
&D_m \eta_- = q_m \eta_- + \im q^\prime_m \gamma^7 \eta_- - \im q_{mn} \gamma^n \eta_+\ .
\end{aligned}
\label{dmeta}
\end{equation}
All $q$ are real with $q_m$ and $q^\prime_m$ transforming in the $\mathbf{3}\oplus\mathbf{\bar 3}$ of $\mathrm{SU}(3)$ while $q_{mn}$ contains the representations $\mathbf{36} = \mathbf{1} \oplus \mathbf{1} \oplus \mathbf{3} \oplus \mathbf{\bar 3} \oplus \mathbf{6} \oplus \mathbf{\bar 6} \oplus \mathbf{8} \oplus \mathbf{8}$. Going to holomorphic indices and using (\ref{jomega}) and (\ref{djdomega}) one can express $q_{mn}$  via the torsion classes \cite{Grana:2004bg} as follows,
\begin{equation}
\begin{aligned}
&q_{\alpha\beta} = - \tfrac\im{16} (\mathcal{W}_3)_{\alpha\bar\gamma\bar\delta} (\Omega_\eta)_\beta^{\bar\gamma\bar\delta} - \tfrac14 (\Omega_\eta)_{\alpha\beta\gamma} \overline{\mathcal{W}}_4^\gamma\ , \\
&q_{\alpha\bar\beta} =  \tfrac14 g_{\alpha\bar\beta} \overline{\mathcal{W}}_1 - \tfrac\im4 (\overline{\mathcal{W}}_2)_{\alpha\bar\beta}\ .
\end{aligned}
\label{qtorsion}
\end{equation}

Equipped with (\ref{dmeta}) and (\ref{qtorsion}) we can now compute the contribution to $W$ due to the torsion. Let us start again with the contribution to the gravitino mass term. It arises from the first term in (\ref{sf}) with the derivative in an internal direction. Inserting the decomposition (\ref{psidecomp}) we find
\begin{equation}
\begin{aligned}
S_{\mathrm{f},\psi} &= - \int_{10} \hat{\bar\psi}_M \Gamma^{MNP} D_N \hat\psi_P \\
&= \tfrac12 \int_4 \bar\psi_\mu \bar\sigma^{[\mu} \sigma^{\nu]}
\bar\psi_\nu \int_6 \eta_-^\dag \gamma^{\bar\alpha} D_{\bar\alpha}
\eta_+ + \mathrm{h.c.}\ ,
\end{aligned}
\label{storsion1}
\end{equation}
where we only kept terms quadratic in the gravitino. Using (\ref{dmeta}), (\ref{qtorsion}) and  (\ref{djdomega}) in the Einstein frame one finds
\begin{equation}
\int_6 \eta_-^\dag \gamma^{\bar\alpha} D_{\bar\alpha} \eta_+ = \tfrac{3\im}{2} \int_{\hat Y} \mathcal{W}_1 = \tfrac\im4 \int_{\hat Y} \Omega_\eta \wedge d\tilde J \ .
\label{propdint}
\end{equation}
Performing the Weyl rescaling according to (\ref{weylresc}), (\ref{resc1}) and (\ref{resc2}) and using (\ref{absomega}) and (\ref{frames}) yields
\begin{equation}
S_{\mathrm{f},\psi} = - \int_4 \bar\psi_\mu \bar\sigma^{\mu\nu} \bar\psi_\nu e^{\frac K2} W_T + \mathrm{h.c.}\ ,
\end{equation}
with
\begin{equation}
W_T = \tfrac\im4 \int_{\hat Y} \Omega \wedge dJ\ .
\label{wtorsion}
\end{equation}
Together with the contribution from $H_3$-flux computed in (\ref{wh}) this yields a superpotential
\begin{equation}
W = W_H + W_T = \tfrac14 \int_{\hat Y} \Omega \wedge (H + \im dJ)\ .
\label{w}
\end{equation}
Let us stress that in this derivation we did not restrict our analysis to the case of half-flat manifolds. 

As before let us now focus on the gravitino-fermion couplings in order to determine the $F$-term contribution of the torsion. For the fermions in the chiral multiplets they arise from $S_{\mathrm{f},\psi}$ after insertion of (\ref{psidecomp}) if we keep the following terms in the expansion
\begin{equation}
\begin{aligned}
S_{\mathrm{f},\psi} = & - \int_{10} \big[ \hat{\bar \psi}_n \Gamma^{nm\nu} D_m \hat\psi_\nu + \hat{\bar\psi}_\mu \Gamma^{\mu mn} D_m \hat\psi_n \big] \\
=&\ 2\im \int_4 \xi^i \sigma^\mu \bar\psi_\mu \int_6 (\omega_i)_{\alpha\bar\beta} \eta^\dag_- \gamma^{\bar\beta} \gamma^{\alpha\bar\gamma} \im q_{\bar\gamma\delta} \gamma^{\delta} \eta_- \\
& \ {} + 2\im \int_4 \zeta^a \sigma^\mu \bar\psi_\mu \Vert\Omega\Vert^{-1} \int_6 (\rho_a)_{\bar\beta\gamma\delta}
{(\bar\Omega_\eta)_{\bar\epsilon}}^{\gamma\delta} \eta^\dag_- \gamma^{\alpha\bar\beta} \gamma^{\bar\epsilon} \im q_{\alpha\zeta} \gamma^\zeta \eta_- + \textrm{h.c.}\ .
\end{aligned}
\label{sfpsi}
\end{equation}
In the term containing $D_m \psi_n$ we performed an integration by parts, so that we actually compute twice the first term in the first line. This is convenient because then we have to consider the action of the internal derivative $D_m$ exclusively on $\psi_\mu$, which implies that only the derivative of the internal spinors $D_m \eta_\pm$ will be needed and not more complex expressions involving the derivatives of the internal forms. Substituting \eqref{qtorsion} in \eqref{sfpsi} and using results analogous to \eqref{wick} and \eqref{int6} one obtains
\begin{equation}
\begin{aligned}
S_{\mathrm{f},\psi} =& -\im \int_4 \xi^i \sigma^\mu \bar\psi_\mu \Bigl[ - 5\im \int_6 (\omega_i)_{\alpha\bar\beta} g^{\alpha\bar\beta} \mathcal{W}_1 + \int_6 (\omega_i)_{\alpha\bar\beta} (\mathcal{W}_2)_{\delta\bar\gamma} g^{\alpha\bar\gamma} g^{\delta\bar\beta} \Bigl] \\
&- \im \int_4 \zeta^a \sigma^\mu \bar\psi_\mu \Vert\Omega\Vert^{-1}
\im \int_{\hat Y} \rho_a \wedge \mathcal{W}_3 + \textrm{h.c.}\ . \\
\end{aligned}
\label{storsion3}
\end{equation}
Using \eqref{omegagcst} we can write
\begin{equation}
\begin{aligned}
-\im\int_6 (\omega_i)_{\alpha\bar\beta} g^{\alpha\bar\beta} \mathcal{W}_1 &= \frac{\tilde{\mathcal K}_i} {2\tilde{\mathcal K}} \int_6 \mathcal{W}_1 = \frac{\tilde{\mathcal{K}}_i}{12\tilde{\mathcal{K}}} \int_{\hat Y} \mathcal{W}_1 \tilde J \wedge \tilde J \wedge \tilde J \\
& = \frac{\tilde{\mathcal{K}}_i}{12\tilde{\mathcal{K}}} \int_{\hat Y}\ \Omega \wedge d\tilde J\ .
\label{torel1}
\end{aligned}
\end{equation}
In the last step, $\mathcal{W}_1 \tilde J \wedge \tilde J \wedge
\tilde J = d\Omega \wedge \tilde J$ was used, which is a consequence
of \eqref{djdomega} in the string frame together with $\mathcal{W}_2 \wedge \tilde J \wedge \tilde J = 0$. Analogously we compute
\begin{equation}
\begin{aligned}
\int_6 (\omega_i)_{\alpha\bar\beta} (\mathcal{W}_2)_{\delta\bar\gamma} g^{\alpha\bar\gamma} g^{\delta\bar\beta} &= \int_{\hat Y} \mathcal{W}_2 \wedge \tilde J \wedge \omega_i \\
&= \int_{\hat Y} d\Omega_\eta \wedge \omega_i - \int_{\hat Y} \mathcal{W}_1 \tilde J \wedge \tilde J \wedge \omega_i \\
&= \int_{\hat Y} \Omega_\eta \wedge d\omega_i - \frac{\tilde{\mathcal{K}}_i}{\tilde{6\mathcal{K}}} \int_{\hat Y} \Omega_\eta \wedge d\tilde J\ .
\end{aligned}
\label{torel2}
\end{equation}
In going from the first to the second line, $\mathcal{W}_2 \wedge \tilde J \wedge \omega_i = d\Omega\wedge \omega_i - \mathcal{W}_1 \tilde J \wedge \omega_i + \ldots$ was used, which also follows from \eqref{djdomega} in the string frame. In the last step, $\mathcal{W}_1 \tilde J \wedge \tilde J \wedge\ \omega_i$ was substituted by twice the expression \eqref{torel1}. Finally, it can also be seen from \eqref{djdomega} that
\begin{equation}
\begin{aligned}
\int_{\hat Y} \rho_a \wedge \mathcal{W}_3 &= \int_{\hat Y} \rho_a \wedge d\tilde{J}.
\end{aligned}
\label{torel3}
\end{equation}
Inserting \eqref{torel1}, \eqref{torel2} and \eqref{torel3} into (\ref{storsion3}) and performing the Weyl rescaling we obtain
\begin{equation}
\begin{aligned}
S_{\mathrm{f},\psi} = &-\tfrac \im{4\sqrt2} \int_4 \xi^i \sigma^\mu  \bar\psi_\mu e^{\frac K2} \Bigl[ - \frac{\im\mathcal{K}_i}{24\mathcal{K}} \int_{\hat Y} \Omega \wedge \im dJ + \int_{\hat Y} \Omega \wedge \omega_i \Bigl] \\
& {} - \tfrac \im{4\sqrt2} \int_4 \zeta^a \sigma^\mu \bar\psi_\mu e^{\frac K2} \int_{\hat Y} \rho_a \wedge \im dJ + \textrm{h.c.}\ .
\end{aligned}
\label{storsion2}
\end{equation}
We also have a contribution arising from inserting the gravitino shift (\ref{gravshift}) into (\ref{storsion1}). Adding this contribution to (\ref{storsion2}) leads to
\begin{equation}
\begin{aligned}
S_{\mathrm{f},\psi} =& -\tfrac \im{4\sqrt2} \int_4 \xi^i \sigma^\mu \bar\psi_\mu \, e^{\frac K2} \Bigl[ \frac{\im\mathcal{K}_i}{3\mathcal{K}} \int_{\hat Y} \Omega \wedge \im dJ + \int_{\hat Y} \Omega \wedge \omega_i \Bigl]  \\
& {} - \tfrac \im{4\sqrt2} \int_4 \zeta^a \sigma^\mu \bar\psi_\mu e^{\frac K2} \, \int_{\hat Y} \rho_a \wedge \im dJ + \textrm{h.c.}\ .
\end{aligned}
\label{storsion5}
\end{equation}

Combining this result with the $H_3$-flux contribution as obtained in (\ref{sh1}) and (\ref{shlambda}) yields
\begin{equation}
\begin{aligned}
S_{F\mathrm{-terms}} =& -\tfrac \im{4\sqrt2} \int_4 \xi^i \sigma^\mu  \bar\psi_\mu\, e^{\frac K2} \Bigl[ \frac{\im\mathcal{K}_i}{4\mathcal{K}} \int_{\hat Y} \Omega \wedge \big( H + \tfrac43 \im dJ \big) + \int_{\hat Y} \Omega \wedge \omega_i \Bigl]  \\
& {} - \tfrac \im{4\sqrt2} \int_4 \zeta^a \sigma^\mu \bar\psi_\mu\, e^{\frac K2} \int_{\hat Y} \rho_a \wedge (H + \im dJ) \\
& {} - \tfrac\im{4\sqrt2} \int_4 \lambda \sigma^\mu \bar\psi_\mu\, e^{\frac K2} (-e^{2\phi^{(4)}}) \int_{\hat Y} \Omega \wedge H + \textrm{h.c.}\ .
\end{aligned}
\label{shtorsion}
\end{equation}
However, this is not yet in the standard supergravity form since  the gravitino-dilatino coupling received no contribution from the torsion. This can be remedied by the following redefinitions
\begin{equation}
\begin{aligned}
\xi^i \to \xi^i - \frac{v^i\, \mathcal{K}_j \xi^j}{12\mathcal{K}}  + \im\, v^i\,  e^{2\phi^{(4)}} \lambda \ , \qquad \lambda \to - \tfrac12 \lambda - \im\, e^{-2\phi^{(4)}} \, \frac{\mathcal{K}_j \xi^j}{8\mathcal{K}}\ .
\end{aligned}
\label{fredef}
\end{equation}
One can show that these transformations leave the kinetic terms (\ref{sreduc3}) and the sum of (\ref{sh1}) and (\ref{shlambda}) (i.e. the $H_3$-flux-dependent part of the $F$-terms) unchanged. Inserting \eqref{fredef} into (\ref{shtorsion}) we obtain
\begin{equation}
\begin{aligned}
S_{F\mathrm{-terms}} = -\tfrac \im{\sqrt2} \int_4 \big[ \xi^i \sigma^\mu & \bar\psi_\mu e^{\frac K2} D_i W + \zeta^a \sigma^\mu \bar\psi_\mu e^{\frac K2} D_a W + \lambda \sigma^\mu \bar\psi_\mu e^{\frac K2} D_S W \big] + \textrm{h.c.}\ ,
\end{aligned}
\end{equation}
where
\begin{equation}
\begin{aligned}
D_i W & = \tfrac14 \Bigl[ \frac{\im\mathcal{K}_i}{4\mathcal{K}} \int_{\hat Y} \Omega \wedge (H + \im dJ) + \int_{\hat Y} \Omega \wedge d\omega_i \Bigl], \\
D_a W & = \tfrac14 \int_{\hat Y} \rho_a \wedge (H + \im dJ), \\
D_S W & = - e^{2\phi^{(4)}} W\ ,
\end{aligned}
\label{dw}
\end{equation}
with $W$ given in (\ref{w}). This establishes the consistency with $N=1$ supergravity.

So far we computed the gravitino mass terms and the mixing of the gravitino with the moduli fermions. In principle we could also compute the mass terms of the matter fermions and their mixing with the gravitino. Such terms correspond to quadratic terms in $W$ which are of the form $m_{ai}(z,t) A^iA^a$, where contraction of gauge indices is implicit. As already shown in \cite{Gurrieri:2007jg} $m_{ai}$ is non-zero for non-vanishing flux and/or torsion. In our formalism such terms do indeed appear from reducing the last term in both \eqref{sf} and \eqref{sint}. However, in order to extract the precise contribution to the superpotential it is necessary to disentangle two different pieces. More precisely, the reduction leads to terms in the effective action proportional to 
 \cite{Wess:1992cp}
\begin{equation}
\chi^i\chi^a\, \Bigl[\frac{\partial^2 W}{\partial A^i\partial A^a} + \frac{\partial^2
K}{\partial A^i\partial A^a}\, W \Bigl]\Bigl|_{A^i=A^a=0} \ .
\end{equation}
$m_{ai}$ is determined by the first term in this expression but for its extraction  we would also need to know the quadratic contribution to the K\"ahler potential which governs the second term. This computation is beyond the scope of this paper and we refer the reader to ref.~\cite{Gurrieri:2007jg} instead.

\subsection{$D$-terms}\label{sec-dterm}

Finally we compute the $D$-terms in the effective action. As can be seen from (\ref{appsint}) in the fermionic action they appear in the coupling of the gravitino to the gaugino. This contribution to the action is coming from the reduction of the similar coupling of the ten-dimensional gravitino $\hat\Psi_M$ to the ten-dimensional gaugino $\hat\chi$ in (\ref{sint}). Performing the reduction of the relevant term is straightforward and leads to
\begin{equation}
\begin{aligned}
S_{\mathrm{int}} &= \tfrac12 \int_{10} e^{-\frac{\hat\phi}4} {\hat F}^{\hat A}_{MN} {\hat{\bar \chi}}^{\hat A} \Gamma^Q \Gamma^{MN} {\hat \psi}_Q \\
&= - \im \int_4 \psi_\mu\sigma^\mu \bar\chi^A e^{-\frac\phi4} \int_6 F^A_{\alpha\bar\beta} g^{\alpha\bar\beta} + \mathrm{h.c.} +\ldots\\
&= \int_4 \psi_\mu \sigma^\mu \bar\chi^x e^{-\frac\phi4} \int F^x \wedge \ast \tilde J + \mathrm{h.c.}+ \ldots\ ,
\end{aligned}
\end{equation}
where the terms hidden in $\ldots$ do not contribute to the $D$-term. In the last step we used
\begin{equation}
\int_6 F^A_{\alpha\bar\beta} g^{\alpha\bar\beta} = \frac \im 2 \int F^A \wedge \tilde J \wedge \tilde J = \im \int F^A \wedge \ast \tilde J\ .
\label{fj}
\end{equation}
After performing the Weyl rescaling (\ref{weylresc}), (\ref{resc1}) and the redefinitions (\ref{resc2}) we obtain
\begin{equation}
S_{\mathrm{int}} = \frac12 \int_4 \psi_\mu \sigma^\mu \bar\chi^A \mathcal{K}^{-1} \int F^A \wedge \ast J\ .
\end{equation}
Comparing with (\ref{appsint}) and recalling (\ref{gaugekineticf}) we conclude that
\begin{equation}
\mathcal{D}^A = - e^{2\phi^{(4)}} \mathcal{K}^{-1} \int F^A \wedge \ast J\ .
\label{dterm}
\end{equation}

In checking this expression we argue that \cite{Gurrieri:2007jg}
\begin{equation}
F^A_{\alpha\bar\beta} T_A = [A_\alpha, A_{\bar\beta}]\ ,
\end{equation}
where in the r.h.s. only contributions along the algebra of the
unbroken gauge group $G$ are taken into account. $T_A$ are the
generators of $G$ and from \eqref{adecomp} and \eqref{arescal} it follows 
\begin{equation}
A_{\alpha} = \tfrac12 \Vert\Omega\Vert^{-\frac13} A^{iP} {(\omega_i)_{\alpha}}^{\beta} T_{\beta P} + \tfrac12 {\Vert\Omega\Vert^{\frac13}} \frac1{\Vert\Omega\Vert^2} A^{a\bar Q} (\bar\rho_a)_{\alpha\bar\gamma\bar\delta} \Omega^{\bar\beta\bar\gamma\bar\delta} {\bar T}_{\bar\beta\bar Q}\ .
\end{equation}
If we take the case of the standard embedding \eqref{standembed} then $T_{\alpha P}$ and $\bar T_{\bar \beta \bar Q}$ will be the generators of the last two factors in the decomposition \eqref{adjdecomp} respectively, while the $T^A$ will span the second factor, the adjoint of $E_6$. It can be seen from the algebra of $E_8$ that in this case
\begin{equation}
[T_{\alpha P},\bar T_{\bar\beta \bar Q}] = \tilde g_{\alpha\bar\beta} t^A_{P\bar Q} T_A + \ldots\ ,
\end{equation}
where the dots refer to generators outside the algebra of $E_6$ and the ${t_{A,P}}^Q$ denote the generators of $E_6$ in the $\mathbf{27}$ representation.

Now we compute
\begin{equation}
\begin{aligned}
&\int_{\hat Y} F^A T_A \wedge \ast J = e^\phi \int_{\hat Y} F^A T_A \wedge \ast \tilde J = e^\phi \int_6 \tilde g^{\alpha\bar\beta} F^A_{\alpha\bar\beta} T_A \\
&\qquad = - \Vert\Omega\Vert^{-\frac23} e^\phi \int_{\hat Y} \omega_i \wedge \ast \omega_j A^{iP} \bar A^{j\bar R} t^A_{P\bar R} T_A 
 - \Vert\Omega\Vert^{\frac23} e^\phi \frac{i\int_{\hat Y} \rho_a \wedge \bar\rho_b} {\Omega\wedge \bar\Omega} \bar A^{aP} A^{b\bar R} t^A_{P\bar R} T_A \\
&\qquad= - \Vert\Omega\Vert^{-\frac23} e^{\frac\phi2} \mathcal{K} g_{ij} A^{iP} \bar A^{j\bar R} t^A_{P\bar R} T_A - \Vert\Omega\Vert^{\frac23} e^{-\frac\phi2} \mathcal{K} g_{a\bar b} \bar A^{aP} A^{b\bar R} t^A_{P\bar R} T_A\ .
\end{aligned}
\end{equation}
Using \eqref{dterm} and \eqref{Z} yields
\begin{equation}
\begin{aligned}
\mathcal{D}^A =& Z_{ij} A^{iP} \bar A^{j\bar R} t^A_{P\bar R} + Z_{a\bar b} \bar A^{aP} A^{b\bar R} t^A_{P\bar R} \\
=& \frac{\partial K_{\mathrm{gauge}}}{\partial A^{iP}} {(t^A)^P}_Q A^{iQ} + \frac{\partial K_{\mathrm{gauge}}}{\partial A^{a\bar P}} {(t^A)^{\bar P}}_{\bar Q} A^{a\bar Q}
\end{aligned}
\end{equation}
for a matter field K\"ahler potential given by
\begin{equation}
K_{\mathrm{gauge}} = Z_{ij} \bar A^i_P A^{jP} + Z_{a\bar b} \bar A^{aQ} A^b_Q\ .
\end{equation}
This, once more, is consistent with $N = 1$ supergravity \cite{Wess:1992cp}.

\section{Supersymmetry transformations}\label{sec-susy}

For completeness let us also compute the fermionic supersymmetry transformation in four space-time dimensions 
from a Kaluza-Klein reduction of the ten-dimensional transformations. This in fact yields an independent computation of $W$ and $K$.

The form of these transformations for a generic $N = 1$ theory in four-dimensions is given in \cite{Wess:1992cp} and repeated in our conventions in eq.~(\ref{4dsusy}). From these equations we see that the gravitino supersymmetry transformation gives directly the superpotential, in analogy to the gravitino mass term in the action, while the transformation of the chiral fermions gives the derivatives of the superpotential with respect to the corresponding scalar superpartners or moduli in analogy to the $F$-terms as well as the
$D$-terms. Let us start with gravitino.

The supersymmetry transformation of the ten-dimensional gravitino is given by \cite{Romans:1985xd}
\begin{equation}
\delta\hat\Psi_M = D_M \hat\epsilon + \tfrac1{96} e^{-\frac{\hat\phi}2} H_{NPQ} \big[ {\Gamma_M}^{NPQ} - 9 \delta_M^N \Gamma^{PQ} \big] \hat\epsilon\ ,
\label{vargrav10d}
\end{equation}
which implies
\begin{equation}
\delta\hat\Psi_\mu = D_\mu \hat\epsilon + \tfrac1{96} e^{-\frac{\hat\phi}2} {\hat H}_{mnp} \Gamma_\mu \Gamma^{mnp} \hat\epsilon\ .
\label{vargrav10d0}
\end{equation}
However, the correct four-dimensional gravitino is only obtained after the shift given in (\ref{gravshift}) which can be interpreted at the level of the ten-dimensional gravitino as follows \cite{Grana:2005ny}
\begin{equation}
\delta\hat\Psi_\mu^\prime \equiv \delta\hat\Psi_\mu + \tfrac12 \Gamma_\mu \Gamma^m \delta \hat\Psi_m\ .
\label{gravshift10d}
\end{equation}
Also from (\ref{vargrav10d}) we compute
\begin{equation}
\begin{aligned}
\Gamma^m \hat\Psi_m & = \Gamma^m D_m \hat\epsilon + \tfrac1{96} e^{-\frac{\hat\phi}2} H_{npq} (\Gamma^m {\Gamma_m}^{npq} - 9 \Gamma^n \Gamma^{pq} ) \hat\epsilon \\
& = \Gamma^m D_m\hat\epsilon - \tfrac1{16} e^{-\frac{\hat\phi}2} {\hat H}_{mnp} \Gamma^{mnp} \hat\epsilon\ ,
\label{vargravint}
\end{aligned}
\end{equation}
where $\Gamma^m {\Gamma_m}^{npq} = 3 \Gamma^{npq}$ was used. After substitution of (\ref{vargrav10d0}) and (\ref{vargravint}) into (\ref{gravshift10d}) we obtain
\begin{equation}
\delta\hat\Psi_\mu^\prime = D_\mu \hat\epsilon + \frac12\Gamma_\mu\Gamma^m D_m \hat\epsilon - \frac1{48}
e^{-\frac{\hat\phi}2} {\hat H}_{mnp} \Gamma_\mu \Gamma^{mnp}
\hat\epsilon\ .
\end{equation}
Inserting (\ref{psidecomp}) and acting with the projector $\tilde{\mathcal K}^{-1} \int_6 {\bf 1}\otimes\eta_-^\dag$ yields (omitting the prime)
\begin{equation}
\begin{aligned}
\delta\psi_\mu = & D_\mu \epsilon + \tfrac \im2 \sigma_\mu \bar\epsilon {\tilde{\mathcal{K}}}^{-1} \int_6 \eta^\dag_- \gamma^{\bar\alpha} D_{\bar\alpha} \eta_+ - \tfrac \im{48} e^{-\frac\phi2} \sigma_\mu \bar\epsilon {\tilde{\mathcal{K}}}^{-1} \int_6 H_{mnp} \eta^\dag_- \gamma^{mnp} \eta_+ \nonumber\\
=& D_\mu \epsilon - \tfrac \im8 e^{-\frac\phi2} \sigma_\mu \bar\epsilon {\tilde{\mathcal{K}}}^{-1}  \Vert\Omega\Vert^{-1} \int_{\hat Y} \Omega \wedge (H + idJ)\ ,
\end{aligned}
\label{notshiftgrav}
\end{equation}
where in the second step (\ref{propdint}) was used. After performing the Weyl rescaling we obtain indeed
\begin{equation}
\delta\psi_\mu = D_\mu \epsilon + \tfrac\im2 \sigma_\mu \bar\epsilon e^{\frac K2} W\ ,
\end{equation}
with $W$ given in (\ref{w}). This shows the agreement with (\ref{4dsusy}).

Let us now turn to the supersymmetry transformations of  the chiral fermions $\xi^i$. In order to do so we first compute
\begin{equation}
(\gamma^5 \otimes \eta_- \eta_-^\dag)\, \Gamma^m \delta\hat\Psi_m\ .
\label{projh11}
\end{equation}
Inserting (\ref{psidecomp}) and  using (\ref{vargravint}) we obtain
\begin{equation}
\begin{aligned}
\delta\bar\xi^i \otimes (\omega_i)_{\alpha\bar\beta}\, \eta_- \eta^\dag_- \gamma^{\bar\beta} \gamma^\alpha \eta_- & = \bar\epsilon \otimes \eta_- \eta^\dag_- \gamma^{\bar\alpha} D_{\bar\alpha} \eta_+ + \tfrac1{16} e^{-\frac\phi2} \bar\epsilon \otimes H_{mnp} \eta_- \eta^\dag_- \gamma^{mnp} \eta_+ \ ,\\
2 \delta\bar\xi^i \otimes (\omega_i)_{\alpha\bar\beta} g^{\alpha\bar\beta} \eta_- & = \tfrac{3\im}2 \bar\epsilon \otimes \mathcal{W}_1 \eta_- + \tfrac1{16} e^{-\frac\phi2} \bar\epsilon \otimes \im H_{mnp} {(\Omega_\eta)}^{mnp}\eta_-\ .
\end{aligned}
\end{equation}
Using the same projector as above we obtain
\begin{equation}
\tilde{\mathcal{K}}_i \delta \bar\xi^i = \tfrac i4\, \bar\epsilon e^{-\frac\phi2}\, \Vert\Omega\Vert^{-1} \int_{\hat Y} \Omega \wedge ( \tfrac32 H + \im dJ )\ ,
\end{equation}
which after the Weyl rescaling reads
\begin{equation}
\mathcal{K}_i \delta \bar\xi^i = \tfrac{\im}{\sqrt2}\, \bar\epsilon e^{\frac K2}  \mathcal{K} \int_{\hat Y} \Omega \wedge ( 3 H + 2 \im dJ )\ .
\label{deltaxi0}
\end{equation}
However, this is not yet in the desired form dictated by (\ref{4dsusy}) since the mixing with the dilatino has not been taken into account yet.

For the ten-dimensional dilatino, the supersymmetry transformation is given by \cite{Romans:1985xd}
\begin{equation}
\delta\hat\lambda = \tfrac{\sqrt2}{48} e^{-\frac{\hat\phi}2} {\hat H}_{MNP} \Gamma^{MNP} \hat\epsilon\ ,
\end{equation}
which after applying the  decomposition \eqref{chilambdadecomp} and the projection leads to
\begin{equation}
\delta\bar\lambda = - \tfrac{\sqrt2}{8}\, \bar\epsilon e^{-\frac\phi2}\, \Vert\Omega\Vert^{-1} \int_{\hat Y} \Omega\wedge H\ .
\end{equation}
After applying the Weyl rescaling this reads
\begin{equation}
\delta\bar\lambda = - \tfrac{\sqrt2}{8}\, \bar\epsilon e^{\frac K2} e^{-2\phi^{(4)}}\int_{\hat Y} \Omega\wedge H\ .
\label{vardil}
\end{equation}
As in the computation of the $F$-terms there is no torsion contribution in this transformation as the mixing with $\xi^i$ has not been taken into account.

Going to new field variables as dictated by the field redefinition (\ref{fredef}) we obtain
\begin{equation}
\begin{aligned}
\mathcal{K}_i \delta \bar\xi^i = & \tfrac\im{\sqrt2}\, \bar\epsilon e^{\frac K2} \im \mathcal{K} \int_{\hat Y} \Omega \wedge ( 3 H + \im dJ ) \ ,\\
\delta\bar\lambda = & - \tfrac{\sqrt2}{8}\,  \bar\epsilon e^{\frac K2} e^{-2\phi^{(4)}} \int_{\hat Y} \Omega \wedge (H + idJ)\ .
\end{aligned}
\end{equation}
From (\ref{dw}) we compute
\begin{equation}
\begin{aligned}
g^{ij} \mathcal{K}_i D_j W & = 3 i\mathcal{K} \int_{\hat Y}\Omega\wedge ( H + \im dJ ) + 2 \mathcal{K} \int_{\hat Y} \Omega \wedge dJ \\
& = \im \mathcal{K} \int_{\hat Y} \Omega \wedge ( 3H + \im dJ )\ ,
\end{aligned}
\end{equation}
and also
\begin{equation}
g^{S\bar S}D_S W = - e^{-2\phi^{(4)}} W\ .
\end{equation}
Therefore we obtain for the supersymmetry transformations
\begin{equation}
\begin{aligned}
\delta \bar\xi^i =& \tfrac1{\sqrt2} \bar\epsilon e^{\frac K2} g^{ij} D_j W\ , \\
\delta\bar\lambda =& \tfrac1{\sqrt2} \bar\epsilon e^{\frac K2} g^{S\bar S} D_S W\ ,
\end{aligned}
\label{deltaxilambda}
\end{equation}
in agreement with (\ref{4dsusy}).

For the supersymmetry transformations of the $\zeta^a$ we evaluate
\begin{equation}
(\gamma^5\otimes \eta_-\eta_-^\dag)\, (\rho_b)_{\bar\alpha\gamma\delta} \bar\Omega^{\beta\gamma\delta} \Gamma^{\bar\alpha} \delta\hat\Psi_{\beta}\ .
\label{projh12}
\end{equation}
Using the decomposition (\ref{psidecomp}), (\ref{metrics}) and (\ref{absomega}) this expression turns into
\begin{equation}
\begin{aligned}
\delta\bar\zeta^a \otimes \eta_- \frac1{\Vert\Omega\Vert^2} (\rho_b)_{\bar\alpha\gamma\delta} (\bar\Omega_\eta)^{\beta\gamma\delta} (\bar\rho_a)_{\beta\bar\epsilon\bar\zeta} {(\Omega_\eta)_\lambda}^{\bar\epsilon\bar\zeta} \eta^\dag_- \gamma^{\bar\alpha} \gamma^\lambda \eta_- & = 8\im \delta\bar\zeta^a \otimes \eta_-\, \frac{\int \rho_b \wedge \bar\rho_a} {\Vert\Omega\Vert^2} \\
& = 8 \tilde{\mathcal{K}} g_{b\bar a} \delta\bar\zeta^a \otimes \eta_-\ .
\label{zetalhs}
\end{aligned}
\end{equation}
On the other hand, using (\ref{vargrav10d}) in (\ref{projh12}) we obtain 
\begin{equation}\label{zetarhs}
\begin{aligned}
\bar\epsilon \otimes \eta_- (\rho_b)_{\bar\alpha\gamma\delta} (\bar\Omega_\eta)^{\beta\gamma\delta}& \Bigl[ \eta^\dag_- \gamma^{\bar\alpha} D_\beta \eta_+ + \tfrac1{96} e^{-\frac\phi2} ( H_{npq} \eta^\dag_- \gamma^{\bar\alpha} {\gamma_\beta}^{npq} \eta_+ - 9 H_{\beta pq} \eta^\dag_- \gamma^{\bar\alpha} \gamma^{pq} \eta_+ ) \Bigl]\\
&= \tfrac12 \bar\epsilon \otimes \eta_- e^{-\frac\phi2} \int \rho_b \wedge (H + idJ)\ ,
\end{aligned}
\end{equation}
where we used (\ref{qtorsion}), (\ref{djdomega}) and
\begin{equation}
H_{npq} \eta^\dag_- \gamma^{\bar\alpha} {\gamma_\beta}^{npq} \eta_+ = -6 \im H^{\bar\alpha\epsilon\zeta} (\Omega_\eta)_{\beta\epsilon\zeta}\ , \qquad H_{\beta pq} \eta^\dag_- \gamma^{\bar\alpha} \gamma^{pq} \eta_+ = \im H_{\beta\bar\epsilon\bar\zeta} (\Omega_\eta)^{\bar\alpha\bar\epsilon\bar\zeta}\ .
\end{equation}
Equating (\ref{zetalhs}) with (\ref{zetarhs}) yields
\begin{equation}
\delta\bar\zeta^a = \tfrac1{16} \bar\epsilon e^{-\frac\phi2} {\tilde{\mathcal{K}}}^{-1} {\Vert\Omega\Vert}^{-1} g^{\bar ab} \int_{\hat Y} \rho_b \wedge (H + \im dJ )\ ,
\end{equation}
which after the Weyl rescaling can be written as
\begin{equation}
\bar\delta\zeta^a = \tfrac1{\sqrt2}\, \bar\epsilon e^{\frac K2} g^{\bar ab} D_b W\ ,
\label{deltazeta}
\end{equation}
with $W$ given again by (\ref{w}).

Finally, we compute the transformation of the gauginos. The ten-dimensional variation is given by
\begin{equation}
\delta\hat{\chi}^{\hat A} = - \tfrac14 e^{\frac{\hat\phi}4} \hat{F}_{MN}^{\hat A} \Gamma^{MN} \hat\epsilon\ .
\end{equation}
The variation of the four-dimensional gaugino $\chi$ is obtained when the index $A$ takes values in the adjoint of the gauge group in four dimensions $G \times E_8$.  Inserting the decomposition of the ten-dimensional gaugino given in (\ref{chilambdadecomp}) leads to
\begin{equation}
\delta\chi^A = \tfrac12 e^{-\frac\phi4} F^A_{\mu\nu} \sigma^{\mu\nu}
\epsilon + e^{-\frac\phi4} \tilde{\mathcal{K}}^{-1} \int_6
F^A_{\alpha\bar\beta} g^{\alpha\bar\beta}\ .
\end{equation}
Substituting (\ref{fj}) and performing the Weyl rescaling (\ref{resc1}) and the redefinitions (\ref{resc2}) we obtain
\begin{equation}
\delta\chi^A = F^A_{\mu\nu} \sigma^{\mu\nu} \epsilon + \im e^{2\phi^{(4)}} \epsilon \mathcal{K}^{-1} \int F^A \wedge \ast J\ .
\label{deltachi}
\end{equation}
Comparing with (\ref{dterm}) we can see the agreement with the supergravity expression (\ref{4dsusy}).

\subsubsection*{Supersymmetry conditions for the vacuum}

With the supersymmetry transformations for the fermions at hand we can discuss the conditions which lead to a supersymmetric background in a flux compactification. In the case of the heterotic string Strominger has shown \cite{Strominger:1986uh} that for a supersymmetric vacuum the background must allow for a non-vanishing torsion. Moreover, the internal manifold has to be complex and the fundamental two-form $J$, the Yang-Mills field strength $F$ and the three-form flux $H_3$ have to satisfy the following conditions\footnote{In \cite{Strominger:1986uh} the condition for the $H_3$ flux includes a factor of $\frac12$, but our normalization for the $H_3$ field strength is twice the one he uses.}
\begin{equation}
J^{\alpha\bar\beta} F_{\alpha\bar\beta} = 0\ , \qquad H_3 = \im (\partial - \bar\partial) J\ .
\end{equation}
Strominger's analysis was made on backgrounds of the form \eqref{backg} which allow for a warp factor $\Delta$. Demanding the vanishing of the gravitino supersymmetry transformation he shows that $\Delta$ is equal to the dilaton. However in our analysis we do not consider any warping, and hence our assumption of a constant dilaton is consistent with Strominger's result in the limit of constant $\Delta$.

On a supersymmetric vacuum the supersymmetry transformations of the fermionic fields vanish, and in particular this must be true for the chiral fermions,
\begin{equation}
\delta \xi^i = \delta \zeta^a = \delta \lambda = \delta \chi^A = 0\ .
\end{equation}
Take for example the transformation of the gaugino, eq.~\eqref{deltachi}. Setting it to zero and considering \eqref{fj} lead to the vanishing of the contraction $F^A_{\alpha\bar\beta} J^{\alpha\bar\beta}$. Hence Strominger's condition on the Yang-Mills field strength is obtained.

The vanishing of the supersymmetry transformation of the dilatino and the $\xi^i$ chiral fermions as given in \eqref{deltaxilambda} requires $D_SW = D_iW = 0$. Considering the expressions for these derivatives given in \eqref{dw} we see that these conditions are indeed equivalent to
\begin{equation}
\begin{aligned}
&\int_{\hat Y} \Omega \wedge (H + \im dJ) = 0\ ,\\
&\int_{\hat Y} \Omega \wedge d \omega_i = \int_{\hat Y} \mathcal{W}_1 J^2 \wedge \omega_i + \int_{\hat Y} \mathcal{W}_2 \wedge J \wedge \omega_i = 0\ .
\end{aligned}
\label{cond}
\end{equation}
From the second of these conditions we conclude that on a supersymmetric background the torsion classes $\mathcal{W}_1$ and $\mathcal{W}_2$ vanish, which is equivalent to saying that the compactification manifold $\hat Y$ is actually complex. 

On the other hand, the first condition in \eqref{cond} tells us that $H_3 + \im dJ$ on a supersymmetric background can only be a sum of $(3,0)$, $(2,1)$ and $(1,2)$ pieces. The $(3,0) + (0,3)$ part of $dJ$ is proportional to $\mathcal{W}_1$ as can be checked from \eqref{djdomega} and therefore vanishes. Now reality of $H_3$ requires that the combination $H_3 + \im dJ$ must be actually of $(2,1) + (1,2)$ type.

Now set to zero the transformation of the $\zeta^a$ chiral fermions eq.~\eqref{deltazeta}. In view of \eqref{dw} it means that
\begin{equation}
\int_{\hat Y} \rho_a \wedge (H + \im dJ) = 0\ ,
\end{equation}
which in turn implies the vanishing of the $(1,2)$ part of $H_3 + \im dJ$. Since the two-form $J$ is a $(1,1)$-form and the NS-flux is $H_3 = H_{(2,1)} + H_{(1,2)}$ one can write the last result for a complex manifold as $H_{(1,2)} = - \im \bar \partial J$, which considering also its conjugate leads to
\begin{equation}
H_3 = \im (\partial - \bar \partial) J
\end{equation}
on a supersymmetric vacuum. So we obtain also the condition on the three-form flux and with it all the supersymmetry conditions for the heterotic string obtained by Strominger \cite{Strominger:1986uh}.

\section{Conclusions}

In this paper we have revisited the compactification of the heterotic string on manifolds $\hat Y$ with $\mathrm{SU}(3)$ structure in presence of background fluxes. We performed the Kaluza-Klein reduction and
computed the K\"ahler potential $K,$ the superpotential $W$ and the $D$-term of the resulting $N = 1$ low-energy supergravity. We applied methods used previously in type II compactifications in order to
identify the light modes \cite{Gurrieri:2002wz,Grana:2005ny}. In particular we projected out all triplets of the $SU(3)$ structure group and expanded in a finite set of differential forms on $\hat Y$. In contrast to the Calabi-Yau case these forms are not necessarily harmonic.

The novel aspect of our paper is that we performed the Kaluza-Klein reduction  in the fermionic sector and determined $K, W$ and $\mathcal{D}$ entirely from fermionic couplings. This has the advantage that $W$ and $\mathcal{D}$ appear linearly and can be read off more easily. As far as we know this procedure has not been carried even for Calabi-Yau compactifications previously. In this respect our paper closes a gap in the existing literature.

Apart from the standard Yukawa couplings for the matter fields we determined the contributions from NS flux and the torsion to superpotential and the $D$-term. The precise structure of the gauge
bundle was left somewhat unspecified in the analysis, as it was not relevant for most of the derivations. However when appropriate we discussed explicitly the case of the standard embedding leading to an unbroken $E_6$ in four dimensions.

Finally we also determined the supersymmetry transformations for the fermionic fields, which is another way to determine $W$ and $\mathcal D$. It also shows the consistency with the expressions for the K\"ahler potential and the superpotential previously derived.

\subsection*{Acknowledgments}

This work is supported by DFG -- The German Science Foundation (DFG) and the European RTN Program MRTN-CT-2004-503369.

We have greatly benefited from conversations and correspondence with Mariana Gra\~na, Thomas Grimm, Olaf Hohm, Paolo Merlatti, Daniel Waldram, Bastiaan Spanjaard and especially Andrei Micu.

\newpage

\noindent {\bf\Large Appendix}

\appendix

\section{Spinors in four and six dimensions}\label{app-conv}

In this appendix we collect the spinor conventions used throughout this paper.

In $D=10$ the $\Gamma$-matrices satisfy the Clifford algebra
\begin{equation}
\{\Gamma^M, \Gamma^N\} = 2 g^{MN}, \qquad M,N = 0,\ldots,9\ .
\end{equation}
One defines \cite{Polchinski:1998rr}
\begin{equation}
\Gamma^{11} = \Gamma^0 \ldots \Gamma^9\ ,
\end{equation}
which has the properties
\begin{equation}
(\Gamma^{11})^2 = 1\ , \qquad \{\Gamma^{11}, \Gamma^M\} = 0\ .
\end{equation}
This implies that the Dirac representation can be split into two Weyl representations,
\begin{equation}
\mathbf{32}_{\mathrm{Dirac}} = \mathbf{16} + \mathbf{16}^\prime\ ,
\end{equation}
with eigenvalues $+1$ and $-1$ under $\Gamma^{11}$, respectively.

In backgrounds of the form (\ref{backg}) the ten-dimensional Lorentz group decomposes as
\begin{equation}
\mathrm{SO}(1,9) \to \mathrm{SO}(1,3) \times \mathrm{SO}(6)\ ,
\end{equation}
implying a decomposition of the spinor representations as
\begin{equation}
\mathbf{16} = (\mathbf{2}, \mathbf{4}) + (\mathbf{\bar 2},
\mathbf{\bar 4})\ . 
\end{equation}
Here $\mathbf{2}$ and $\mathbf{4}$ are the spinor representations of $\mathrm{SO}(1,3)$ and $\mathrm{SO}(6)$, respectively.

In the background (\ref{backg}) the ten-dimensional $\Gamma$-matrices can be chosen block-diagonal as
\begin{equation}
\Gamma^M = (\gamma^\mu \otimes \mathbf{1}, \gamma^5 \otimes \gamma^m)\
,
\qquad \mu = 0, \ldots, 3,\quad  m = 1, \ldots, 6\ ,
\label{gammadecomp}
\end{equation}
where $\gamma^5$ defines the Weyl representations in $D=4$. In this basis, $\Gamma^{11}$ splits as \cite{Polchinski:1998rr} 
\begin{equation}
\Gamma^{11} = -\gamma^5 \otimes \gamma^7\ ,
\end{equation}
where $\gamma^7$ defines the Weyl representations in six dimensions.

Let us now focus on our spinor conventions in four- and six-dimensions.

\subsection{Clifford algebra in four-dimensions}

In $D=4$ we adopt the conventions of \cite{Wess:1992cp}  and choose
\begin{equation}\gamma^\mu = - \im \left( \begin{array}{cc} 0& \sigma^\mu\\ \bar\sigma^\mu& 0 \end{array} \right), \qquad \gamma^5 = \left( \begin{array}{cc} \mathbf{1}& 0\\ 0& -\mathbf{1} \end{array} \right),
\end{equation}
where the $\sigma^\mu$ are the $2\times 2$ Pauli matrices
\begin{equation}
\begin{aligned}
&\sigma^0 = \left( \begin{array}{cc} 1& 0\cr 0& 1 \end{array} \right), \qquad \sigma^1 = \left( \begin{array}{cc} 0& 1\cr 1& 0 \end{array} \right),\\
&\sigma^2 = \left( \begin{array}{cc} 0& -\im\cr \im& 0 \end{array} \right), \qquad \sigma^3 = \left( \begin{array}{cc} 1& 0\cr 0& -1 \end{array} \right),
\end{aligned}
\end{equation}
and $\bar\sigma^0 = \sigma^0$, $\bar\sigma^{1,2,3} = - \sigma^{1,2,3}$.

\subsection{Clifford algebra in six-dimensions}

In six-dimensions the $\gamma$-matrices are chosen hermitian, ${\gamma^m}^\dag = \gamma^m$, and they obey the Clifford algebra
\begin{equation}
\{\gamma^m, \gamma^n\} = 2g^{mn}\ , \qquad m,n = 1,\ldots,6\ .
\end{equation}
The Majorana condition on a spinor $\eta$ reads
\begin{equation}
\eta^\dag = \eta^T C\ ,
\end{equation}
where $C$ is the charge conjugation matrix satisfying
\begin{equation}
C^T = C, \qquad \gamma_m^T = -C \gamma_m C^{-1}\ .
\end{equation}
One can regroup these six $\gamma$-matrices into two sets of anticommuting raising and lowering operators \cite{Polchinski:1998rr}
\begin{equation}
\gamma^\alpha = \frac1{\sqrt2}(\gamma^m + \im \gamma^{m+3})\ , \qquad \gamma^{\bar\alpha} = \frac1{\sqrt2}(\gamma^m - \im \gamma^{m+3})\, \qquad \textrm{for}\quad m = 1,2,3\ ,
\end{equation}
satisfying
\begin{equation}
\{\gamma^\alpha, \gamma^{\bar\beta}\} = 2g^{\alpha\bar\beta}\ , \qquad \{\gamma^\alpha, \gamma^\beta\} = 0 = \{\gamma^{\bar\alpha}, \gamma^{\bar\beta}\}\ .
\label{cliffholo}
\end{equation}
In this basis, the two chiral spinors $\eta_\pm$ are annihilated by $\gamma^\alpha$ and $\gamma^{\bar\alpha}$, respectively, that is
\begin{equation}
\gamma^\alpha \eta_+ = 0\ , \qquad \gamma^{\bar\alpha} \eta_- = 0\ .
\label{annihil}
\end{equation}

\section{$N=1$ supergravity in $D=4$}\label{app-sugra}

In this appendix we recall the couplings of $N=1$ supergravity which we need for comparison in the main text. We use the notation and conventions of ref.~\cite{Wess:1992cp}. Let us denote the components of the vector multiplets by $(A_\mu, \chi)$ while we collectively denote the components of the chiral multiplets by $(\Phi^I, \Xi^I)$, $i=1, \ldots, n_c$. In order to facilitate the comparison with the results in the main text we split the action as follows
\begin{equation}
S = S_\mathrm{b} + S_\mathrm{f} + S_\mathrm{int} + \ldots\ ,
\label{s}
\end{equation}
where we neglect terms which are irrelevant for our analysis. $S_\mathrm{b}$ denotes the bosonic action and is given by\footnote{The shorthand $\int_4$ stands for $\int dx^4 \sqrt{-g_4}$.}
\begin{equation}
S_\mathrm{b} = - \int_4 \Bigl[\tfrac12 R + \tfrac14 \mathrm{Re} f(\Phi) \mathrm{Tr} F_{\mu\nu} F^{\mu\nu} - \tfrac14 \mathrm{Im} f(\Phi) \mathrm{Tr} F \tilde F + g_{I\bar J} \partial_\mu \Phi^I \partial^\mu \bar\Phi^{\bar J} + V \Bigl]\ ,
\label{bosonicaction}
\end{equation}
where $f(\Phi)$ is the holomorphic gauge kinetic function and $g_{I\bar J}$ is the K\"ahler metric
\begin{equation}
g_{I\bar J} = \frac{\partial}{\partial \Phi^I} \frac{\partial}{\partial \bar\Phi^{\bar J}} K(\Phi, \bar\Phi)\ ,
\end{equation}
which is determined by the K\"ahler potential $K(\Phi, \bar\Phi)$. Finally, the scalar potential $V(\Phi, \bar\Phi)$ is given as a function of the superpotential $W(\Phi)$
\begin{equation}
V(\Phi, \bar\Phi) = e^K (D_I W g^{I\bar J} \overline{D}_{\bar J} \overline{W} - 3 \vert W \vert^2) + \tfrac12 (\mathrm{Re} f)^{-1} \mathcal{D}^2\ ,
\label{v}
\end{equation}
where
\begin{equation}
D_I W = \frac{\partial W}{\partial\Phi^I} + \frac{\partial K}{\partial \Phi^I} W
\label{kahlerderiv}
\end{equation}
is the K\"ahler derivative of the superpotential and $\mathcal{D}$ is the $D$-term.

The second term in (\ref{s}) comprises the kinetic terms for the fermions
\begin{equation}
S_\mathrm{f} = - \im \int_4 \Bigl[ \im \epsilon^{\mu\nu\rho\lambda} \bar\psi_\mu \bar\sigma_\nu D_\rho \psi_\lambda + g_{\bar I J} \bar\Xi^{\bar I} \bar\sigma^\mu D_\mu \Xi^J + \mathrm{Re}f \bar\chi \bar\sigma^\mu D_\mu \chi \Bigl]\ ,
\label{fermionicaction}
\end{equation}
where $\psi_\mu$ is the gravitino. Finally, we also need the gravitino mass term, the gravitino-fermion couplings and the Yukawa couplings. They are given by
\begin{equation}
\begin{aligned}
S_\mathrm{int} = &- \int_4 \Bigl[ \bar\psi_\mu \bar\sigma^{\mu\nu}
\psi_\nu\, e^{\frac K2} W + \tfrac\im{\sqrt2} \Xi^{ I} \sigma^\mu \bar
\psi_\mu \, e^{\frac K2} D_I W + \tfrac12 (\mathrm{Re}f) \mathcal{D}\,
\psi_\mu \sigma^\mu \bar\chi \\
&\qquad\quad + \tfrac12 \Xi^{ I} \Xi^{ J} D_I D_J W + \mathrm{h.c.} \Bigl]\ ,
\label{appsint}
\end{aligned}
\end{equation}
where $\bar\sigma^{\mu\nu} = \frac14 \bar\sigma^{[\mu} \sigma^{\nu]}$.

Finally, the supersymmetric transformations of the gravitino and the fermions in the chiral multiplets, excluding higher order terms in fermionic fields, are given by
\begin{equation}
\begin{aligned}
\delta\psi_\mu & = D_\mu \epsilon + \tfrac\im2 \sigma_\mu \bar\epsilon e^{\frac K2} W\ , \\
\delta\bar\Xi^{\bar I} & = \tfrac1{\sqrt2} \bar\epsilon e^{\frac K2} g^{\bar IJ} D_J W\ , \\
\delta\chi & = F_{\mu\nu} \sigma^{\mu\nu} \epsilon - \im \epsilon \mathcal{D}\ .
\end{aligned}
\label{4dsusy}
\end{equation}

\section{The geometry of the scalar manifold in $\mathrm{SU}(3)$ compactifications}\label{app-geom}

In this appendix we collect the results of the geometry of the scalar manifold arising in compactifications on manifolds with $\mathrm{SU}(3)$ structure. For the special case of Calabi-Yau manifolds this geometry coincides with the geometrical moduli space of the deformations of the Calabi-Yau metric \cite{Strominger:1985ks,Candelas:1990pi}. For more general manifolds one can still define metric deformations and a metric on the space of metric deformations. The resulting geometry has been discussed in references \cite{Gurrieri:2002wz,Grana:2005ny,KashaniPoor:2006si} and shown to be a product manifold of the form
\begin{equation}
\mathcal{M} = \mathcal{M}_J \times \mathcal{M}_\Omega\ ,
\end{equation}
where $\mathcal{M}_J$ corresponds to the deformations of $J$ while $\mathcal{M}_\Omega$ corresponds to the deformations of the three-form $\Omega$. $N=1$ supersymmetry constrains this product to be a K\"ahler manifold. However, for compactifications on manifolds with $\mathrm{SU}(3)$ structure each factor is a special K\"ahler geometry in that the K\"ahler potential is a sum of two terms
\begin{equation}
K = K_J + K_\Omega\ ,
\end{equation}
and both K\"ahler potentials can be derived from a holomorphic prepotential. Let us discuss this in more detail.

\subsection{The $\mathcal{M}_J$ component}

The coordinates of $\mathcal{M}_J$ are the scalars $t^i = b^i + \im v^i$ which arise from expanding $B_2 + \im J$ in a set of two-forms $\omega_i$ as done in (\ref{b2}) and (\ref{deltag}). The metric on this space is defined as
\begin{equation}
g_{ij} = \frac1{4\mathcal{K}} \int_{\hat Y} \omega_i \wedge \ast \omega_j, \qquad i,j = 1, \ldots, h^{(1,1)}\ ,
\label{gij}
\end{equation}
where $\mathcal{K}$ is the volume of $\hat Y$,
\begin{equation}
\mathcal{K} = \frac16 \int_{\hat Y} J\wedge J \wedge J\ .
\end{equation}
$\ast\omega_j$ denotes a set of four-forms which are dual to the set of two-forms $\omega_i$. It satisfies  \cite{KashaniPoor:2006si}
\begin{equation}
\ast\omega_i = - J\wedge \omega_i + \frac{\mathcal{K}_i}{4\mathcal{K}} J\wedge J\ .
\end{equation}
Inserted into (\ref{gij}) leads to
\begin{equation}
g_{ij} = - \frac1{4\mathcal{K}} \Bigl ( \mathcal{K}_{ij} - \frac1{4\mathcal{K}} \mathcal{K}_i \mathcal{K}_j \Bigl)\ ,
\label{gkk}
\end{equation}
where we abbreviate
\begin{equation}
\mathcal{K}_i = \int_{\hat Y} \omega_i \wedge J \wedge J\ , \qquad \mathcal{K}_{ij} = \int_{\hat Y} \omega_i \wedge \omega_j \wedge J\ , \qquad \mathcal{K}_{ijk} = \int_{\hat Y} \omega_i \wedge \omega_j \wedge \omega_k\ .
\end{equation}
$g_{ij}$ is a K\"ahler metric of the K\"ahler potential $K = - \mathrm{ln} \mathcal{K}$, i.e.
\begin{equation}
g_{ij} = \partial_{t^i} \partial_{\bar t^j} K\ , \qquad K = - \mathrm{ln} \tfrac16 \int_{\hat Y} J\wedge J \wedge J\ .
\end{equation}
In fact, $g_{ij}$ is even a special K\"ahler metric in that $K$ can be derived from a holomorphic prepotential $\mathcal{F}$ via
\begin{equation}
K = - \mathrm{ln} \left[ X^{\hat i} \bar{\mathcal{F}}_{\hat i} - \bar X^{\hat i} \mathcal{F}_{\hat i} \right]\ , \qquad \mathcal{F}_{\hat i} = \partial_{X^{\hat i}} \mathcal{F}\ , \qquad \hat i = 0, \ldots, h^{(1,1)}\ ,
\end{equation}
for
\begin{equation}
\mathcal{F} = (X^0)^2 \mathcal{K}_{ijk} t^i t^j t^k\ , \qquad t^i \equiv \frac{X^i}{X^0}\ .
\label{cubic}
\end{equation}

\subsection{The $\mathcal{M}_\Omega$ component}

$\mathcal{M}_\Omega$ is spanned by the complex scalars $z^a$ with a metric \cite{Candelas:1990pi,Gurrieri:2002wz,Grana:2005ny, KashaniPoor:2006si}
\begin{equation}
g_{a\bar b} = \frac{\int_{\hat Y} \rho_a \wedge \bar\rho_b}{\int_{\hat Y} \Omega \wedge \bar \Omega}\ ,
\end{equation}
where $\rho_a$ and $\Omega$ are $(2,1)$-forms and the $(3,0)$-form, respectively, satisfying the relations \cite{KashaniPoor:2006si}
\begin{equation}
\frac{\partial\Omega}{\partial z^a} = -\frac{\partial K_{\Omega}}{\partial z^a} \Omega + \rho_a\ .
\label{deromega}
\end{equation}
This metric is special K\"ahler with a K\"ahler potential given by
\begin{equation}
K_\Omega = -\mathrm{ln}\Bigl( \im \int_{\hat Y} \Omega\wedge\bar\Omega \Bigl).
\end{equation}
Furthermore, $\Omega$ can be expanded in terms of a real symplectic basis $(\alpha_A, \beta^B)$ of three-forms,
\begin{equation}
\Omega = Z^A(z) \alpha_A - \mathcal{G}_B(z) \beta^B\ , \qquad A,B = 0, \ldots, h^{(1,2)}\ ,
\label{omegadecomp}
\end{equation}
where $\mathcal{G}_A = \partial_{Z^A} \mathcal{G}$. The $z^a$ used in (\ref{deltag}) are the special coordinates defined as $z^a = Z^a/Z^0$.

The holomorphic three-form $\Omega$ is related to $\Omega_\eta$ defined in (\ref{jomega}) by a scale factor
\begin{equation}
\Omega = \Omega_\eta \Vert \Omega \Vert\ , \qquad \Vert\Omega\Vert = e^{\frac12 \tilde{K}_J - \frac12 K_\Omega}\ ,
\label{absomega}
\end{equation}
where $\tilde K_J = - \mathrm{ln} \tilde{\mathcal{K}}$ depends on the volume $\tilde{\mathcal{K}}$ in the Einstein frame. In terms of the components of $\Omega$ one thus has
\begin{equation}
\Omega^{\alpha\beta\gamma} = \epsilon^{\alpha\beta\gamma} \Vert\Omega\Vert\ .
\label{omegaeps}
\end{equation}
The holomorphic three-form $\Omega$ and the $(2,1)$-form satisfy
\begin{equation}
\ast\Omega = -\im \bar \Omega\ , \qquad \ast\rho_a = \im {\bar\rho}_a\ .
\label{astomega}
\end{equation}

\end{document}